\documentclass[letterpaper,referee]{raa}            

\usepackage{graphicx,times}             
\usepackage{natbib}
\usepackage{amssymb,amsmath}
\bibpunct{(}{)}{;}{a}{}{,}

\usepackage[letterpaper]{geometry}
\usepackage[pagebackref=true]{hyperref}
\hypersetup{colorlinks = true, linkcolor = green, anchorcolor = red, citecolor = blue, filecolor = red, urlcolor = red}

\begin{document}

   \title{Structure and Evolution of Magnetic Fields Associated with Solar Eruptions
}

   \volnopage{Vol.0 (200x) No.0, 000--000}      
   \setcounter{page}{1}          

   \author{Haimin Wang
   \and Chang Liu
   }

   \institute{Space Weather Research Laboratory and Big Bear Solar Observatory, New Jersey Institute of Technology, University Heights, Newark, NJ 07102-1982, USA; {\it haimin.wang@njit.edu}\\
   }

   \date{Received~~2014 month day; accepted~~2014~~month day}

\abstract{This paper reviews the studies of solar photospheric magnetic field evolution in active regions and its relationship to solar flares. It is divided into two topics, the magnetic structure and evolution leading to solar eruptions and the rapid changes of photospheric magnetic field associated with eruptions. For the first topic, we describe the magnetic complexity, new flux emergence, flux cancellation, shear motions, sunspot rotation, and magnetic helicity injection, which may all contribute to the storage and buildup of energy and triggering of solar eruptions.  For the second topic, we concentrate on the observations of rapid and irreversible changes of photospheric magnetic field associated with flares, and the implication on the restructuring of three-dimensional magnetic field. In particular, we emphasize the recent advances in observations of photospheric magnetic field, as state-of-the-art observing facilities (such as Hinode and Solar Dynamic Observatory) become available. The linkage between observations and theories and future prospectives in this research area are also discussed.}


   \authorrunning{H. Wang \& C. Liu}            
   \titlerunning{Magnetic Fields and Solar Flares}  

   \maketitle

%
%
\section{Magnetic Structure and Evolution Leading to Solar Eruptions}           
\label{sect:structure}

One of the greatest scientific discoveries in the last century is the existence of magnetic fields in sunspots \citep{hale1908}. Following this discovery, the detailed structure and polarities of magnetic fields in solar active regions (ARs) have been investigated extensively \citep[e.g.,][]{richardson48}. In the large scale, magnetic field and its evolution play a crucial role in the generation of solar activity cycle. In the short term, the level of complexity of magnetic field can be linked to the productivity of solar flares. It is generally believed that magnetic field provides the energy for the solar energetic events, namely, flares and coronal mass ejections (CMEs). Although the details of energy storage and release have not yet been fully understood, from the observational point of view the frequency and intensity of activities observed in the solar corona correlate well with the size and complexity of the host ARs. Furthermore, the evolution of photospheric magnetic field is coupled to the surface flow field, which provides important information about the energy build-up and flare triggering. In recent years, there have been significant advances in the study of magnetic structure and evolution, owing to the availability of the state-of-the-art observations and the development of advanced data analysis and modeling tools. In this section, we review the current understanding of both the static and dynamic pre-flare conditions. Clearly, the related studies not only reveal the physical mechanisms of flare triggering, but also deepen our knowledge on the forecast of solar eruptions, which are the source of the geomagnetic and particle effects in the near-Earth environment. Therefore, these studies are an important part of the space weather research.

\subsection {Static Pre-flare Condition}

\begin{figure}
   \centering
   \includegraphics[width=11.0cm, angle=0]{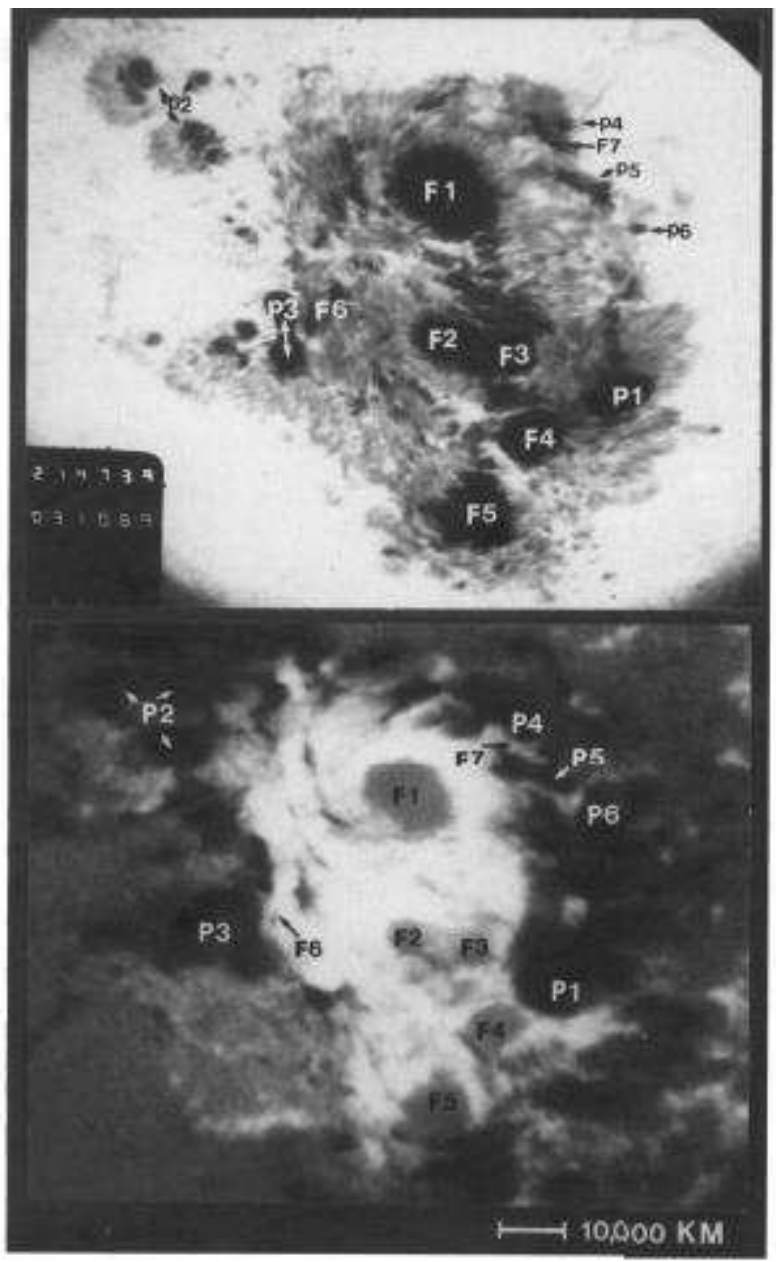}
   \caption{Top: A BBSO He~{\sc i} D3 (a proxy for white light) image of AR 5395 observed on 1989 March 10. This AR produced a number of flares including 10 X-class flares. Bottom: A corresponding LOS magnetogram. This is a typical island $\delta$ sunspot, in which the central sunspots in positive field is surrounded by many patches of negative polarity fields \citep{wang91}.}
   \label{fig1}
   \end{figure}

Significant attention was paid to the classification of the magnetic complexity of sunspots even in earlier years. One of the most used classification scheme is the Mount Wilson classification \citep{hale38}, in which sunspots are categorized into $\alpha$, $\beta$, $\gamma$, and $\delta$ configurations with an increasing magnetic complexity. The $\alpha$ and $\beta$ sunspots have a simple unipolar and bipolar structure, respectively, and these two classes of sunspots have a relatively small chance to produce flares. $\gamma$ spots have mixed polarities, while $\delta$ sunspots are most complicated, with two umbrae of opposite polarities sharing a common penumbra. Most ARs may have a character of combinations of these classifications due to the existence of different activity centers.  \citet{kunzel60} was the first to link the flare productivity with $\delta$ sunspots. A more significant correlation between $\delta$ sunspots and the production of major flares were revealed by \citet{zirin87}, in which 18 years of data from Big Bear Solar Observatory (BBSO) were analyzed to study the development of $\delta$ sunspot and its association with flares. In particular, the authors introduced the term ``island $\delta$ sunspot'', which are compact, elongated $\delta$ sunspots and could be the most flare productive ARs. For example, \citet{wang91} studied the well-known AR 5395 in March of 1989, which produced a number of large flares and caused significant space weather effects. Figure~\ref{fig1} shows the BBSO D3 and magnetograph observations of this AR on 1989 March 10. The dominant polarity of this AR is positive, while many small negative umbrae surround the central positive umbra, forming an extended magnetic polarity inversion line (PIL). Multiple flares were observed to occur in the different sections of this PIL.

Other sunspot classifications are also used to describe the size and complexity of ARs related to flare productivity, and therefore have received attention in the usefulness of solar flare prediction. For example, \citet{mcintosh90} introduced the ``McIntosh'' sunspot classification, which is modified from the earlier Zurich classification \citep{kiepenheuer53}. It includes 60 distinct types of sunspot groups, and has been used as an expert system to predict flares \citep[e.g.,][]{gallagher02,bloomfield12}. The McIntosh classification is currently used by the Space Weather Prediction Center of NOAA.

Before the availability of vector magnetic field data, line-of-sight (LOS) magnetograms alone were used extensively to study solar magnetic fields. One way to analyze the non-potentiality of ARs based on the LOS magnetic field observation is to carry out potential field extrapolations, and compare the structure of the derived three-dimensional (3D) magnetic field with that of the observed corona loops. \citet{schrijver05} classified ARs into flare-active and flare-quiet regions based on the comparison of extrapolated fields with coronal loops observed by the TRACE satellite. They found that in ARs with non-potential coronae, flares occur 2.4 times more frequently and their average X-ray peak brightness is 3.3 times higher than those flares in near-potential regions.

Besides sunspot classifications based on LOS magnetograms, vector magnetograph observations can produce important and useful magnetic parameters for characterizing the non-potentiality of ARs. Several ground-based observatories provide valuable vector magnetograms before such data from space observations became available in recent years. The facilities include BBSO, Marshal Space Flight Center, Huairou Solar Observing Station of National Astronomical Observatories of Chinese Academy of Sciences, and Mees Solar Observatory of University of Hawaii. \citet{wang96} used vector data from Huairou and derived the vertical current as a proxy for non-potentiality. The authors found that the flare activity in AR 6233 was closely associated with the vertical electric current. In an effort to identify an activity-productive indicator, numerous photospheric magnetic properties have been explored \citep[e.g.,][]{schrijver07,jing06,song09,welsch09}. A number of magnetic parameters have been used to predict solar flares, such as surface magnetic free energy \citep{wang96,leka03a,leka03b,falconer06}, the unsigned magnetic fluxes averaged in different ways \citep{schrijver07,barnes06,georgoulis07,jing06}, magnetic shear and magnetic gradient \citep{1984SoPh...91..115H,song09,falconer01,falconer03,2012A&A...543A..49C,li13}, and magnetic energy dissipation \citep{abramenko05}. \citet{zhou09} presented a study of the relationship between magnetic shear and flare shear. In a more recent analysis, \citet{song13} used several parameters to quantify the magnetic complexity in AR 11158 and found a correlation with flares in this region.

The above assessment of magnetic non-potentiality is based on the surface measurement only; however, it is likely that the energy to power flares is stored in the solar corona. As the coronal magnetic field is not directly measured with high resolution and precision, extrapolating magnetic field from the observed photospheric boundary becomes important. Based on LOS magnetograms, potential (current-free) field can be easily derived as done in \citet{schrijver07}. It can be further assumed that magnetic field lines run parallel to electric currents and that the ratio between current and field strength is a constant $\alpha$ (i.e., the so-called linear force-free condition). Quick results of non-potential coronal fields can be obtained this way \citep[e.g.,][]{gary89}, and this $\alpha$ can also be used to evaluate the non-potentiality of ARs. Nevertheless, the assumption of a single force-free parameter is far from the reality.

One of the most advanced coronal magnetic field modeling tools up to date is the non-linear force-free field (NLFFF) extrapolation, in which $\alpha$ can vary among different field lines. \citet{schrijver06} and \citet{metcalf08} summarized and compared various techniques for implementing the NLFFF assumption. It was shown that the method developed by \citet{wiegelmann04} has certain advantages in modeling the 3D coronal field. The extrapolation endeavors have been plagued by the problem that the photospheric magnetic field is not necessarily consistent with the force-free condition. To deal with this \citet{wiegelmann06} preprocessed the vector magnetogram in order to drive the observed non-force-free photospheric data towards the suitable boundary condition in the chromosphere for a force-free extrapolation. The preprocessing routine minimizes a functional that includes two terms respectively corresponding to the force-free and torque-free conditions, one term controlling how close the preprocessed data are compared to the original magnetogram (noise-level), and one term controlling the smoothing. This preprocessing method removes the net force and torque from the photosphere boundary, and hence provides an improved input for the subsequent NLFFF extrapolation \citep{metcalf08}.

It is fortunate that the Helioseismic and Magnetic Imager (HMI) on board Solar Dynamics Observatory (SDO) that was launched in 2010 provides full-disk vector magnetograms, overcoming the limitation of field of views (FOVs) in previous observations. Meanwhile, substantial progress has been made over the past few years in improving the boundary data treatment and the NLFFF extrapolation method. For example, a new version of Wiegelmann's extrapolation code has been developed recently to take the curvature of the Sun's surface into account, so that extrapolations can be applied to large areas \citep{tadesse13}. In addition, the new version also considers measurement errors in photospheric vector magnetograms and keeps a balance between the force-free constraint and the deviation from the photospheric field measurements \citep{wiegelmann12}. These improvements allow a more accurate evaluation of magnetic free energy in 3D as well as magnetic helicity density \citep[e.g.,][]{su14}.

\subsection {Dynamic Pre-flare Condition}

The above discussions of magnetic non-potentiality are mainly based on individual magnetograms, which only provide snap-shots of magnetic conditions. It is well known that magnetic field evolution plays an even more important role in the energy build-up and flare triggering. In this section, we review the dynamic pre-flare conditions, which include new flux emergence, shear and converging flows, sunspot rotation, and magnetic helicity injection.  We start with a summary in deriving the flow field, a key parameter closely associated with the dynamic properties of magnetic field. Time sequence of magnetograms are typically analyzed to understand the dynamic pre-flare conditions.

\subsubsection {Tracking Flows of Solar Active Regions}

There are several plasma velocity inversion methods as summarized and compared by \citet{welsch07}. The most used method in earlier years is based on local correlation tracking developed by \citet{1988ApJ...333..427N}. Its first implementation on time sequence of magnetograms was carried out by \citet{2001ApJ...560L..95C}. Considering the interaction between flow and magnetic fields in terms of induction equations, \citet{2002ApJ...577..501K} and \citet{2003SoPh..215..203D} carefully treated the difference between the horizontal plasma velocity and the apparent velocity due to flux transport. Since then, a number of other methods have been developed, including Fourier Local Correlation Tracking \citep[FLCT;][]{2004ApJ...610.1148W}, Inductive Local Correlation Tracking \citep[ILCT;][]{2004ApJ...610.1148W}, Minimum Energy Fit \citep[MEF;][]{2004ApJ...612.1181L}, Minimum Structure Method \citep[MSM;][]{2006ApJ...636..475G}, and Differential Affine Velocity Estimator \citep[DAVE;][]{2005ApJ...632L..53S,2006ApJ...646.1358S}.

It is notable that the first time derivation of both the vertical and horizontal velocity fields was achieved by the DAVE for vector magnetograms \citep[DAVE4VM;][]{2008ApJ...683.1134S}. This method explicitly incorporates the horizontal magnetic field that is necessary for the description of the vertical flow, and hence makes great progress in estimating the vector velocity field on the photosphere. The performance of DAVE4VM was evaluated using the same synthetic data as used by \citet{welsch07}. It was demonstrated that DAVE4VM substantially outperforms DAVE and is roughly comparable to the MEF method, which was deemed the overall best performing algorithm \citep{welsch07}. Furthermore, DAVE4VM is more accurate than MEF in the estimates of the normal component of perpendicular plasma velocity $v_n$, which is crucial to diagnosing flux emergence and calculating emergence-helicity flux \citep{2008ApJ...683.1134S}. Recently, DAVE4VM has started to be applied to the SDO data \citep[e.g.,][]{2013SoPh..287..279L}. \citet{2008ApJ...689..593C} introduced another improved method, the Nonlinear Affine Velocity Estimator (NAVE), and demonstrated that NAVE is more consistent with the simulated data. Obviously, these developments will facilitate the photospheric flow tracking using advanced data sets from space missions (e.g., Hinode and SDO) as well as high-resolution observations from ground-based facilities.

\subsubsection {Magnetic Flux Emergence and Cancelation}

\begin{figure}
   \centering
   \includegraphics[width=12.0cm, angle=0]{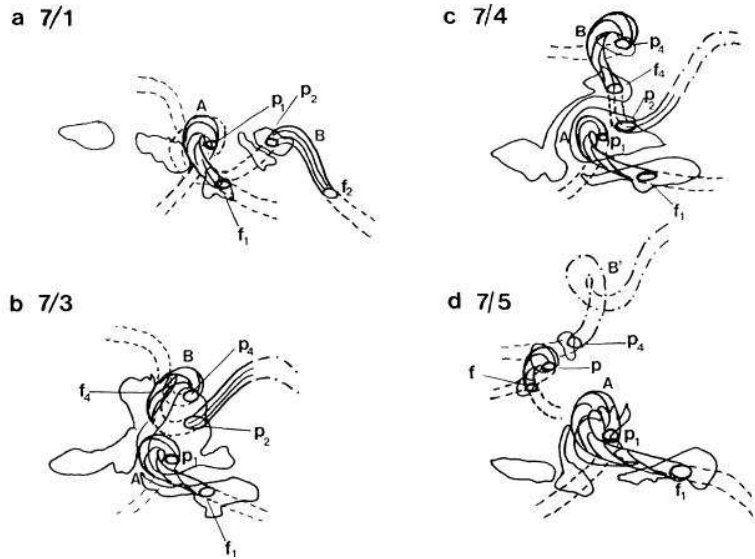}
   \caption{Topological structure of a sunspot as derived from time sequence of photospheric observations. Two magnetic ropes emerge to form a complicated $\delta$ configuration from 1974 July 1 to 5 in AR McMath 13043 \citep{1991SoPh..136..133T}, courtesy Solar Physics.}
   \label{f2}
\end{figure}

\begin{figure}
   \centering
   \includegraphics[width=11.0cm, angle=0]{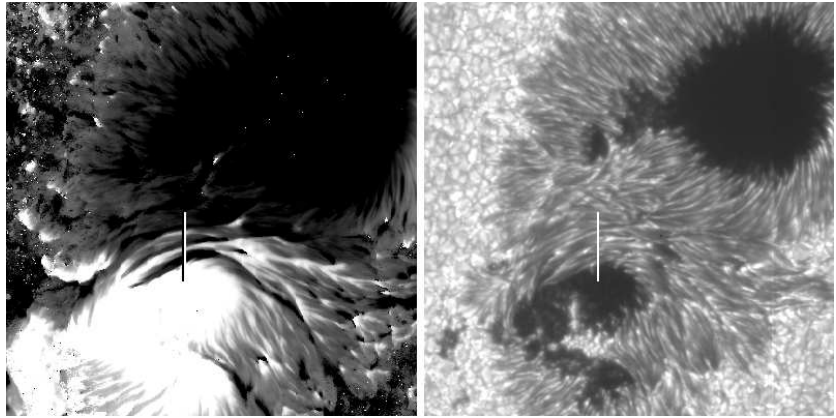}
   \caption{A LOS magnetogram (left) that was obtained by Hinode/SOT and the corresponding G-band image (right) at the peak development of magnetic channels around 12:00~UT on 2006 December 13. The vertical line marks the most prominent part of the magnetic channels. The
FOV is 48\arcsec~$\times$~48\arcsec~\citep{2008ApJ...687..658W}.}
   \label{f3}
\end{figure}

The importance of emerging flux regions (EFRs) leading to solar eruptions was noted many years ago \citep[e.g.,][]{1973SoPh...32..173Z}. \citet{1991SoPh..136..133T} derived the complex subsurface magnetic structure in a flare-productive $\delta$ sunspot group. This sunspot group produced many large flares and showed unusually fast changes of magnetic structure. The comprehensive observations provided an excellent opportunity to find the relationship between the flare occurrence and evolution of the magnetic configuration. The author inferred magnetic topological structure of the region in the form of a long-winding twisted rope with a number of twisted knots. The unusual evolution of this $\delta$ group as observed in the  photosphere was explained by consecutive emergences of a single system through the observable surface. Figure~\ref{f2} shows the cartoon of Tanaka, which demonstrates the topological structure derived from time sequence of photospheric observations. Similar methodology was used by \citet{1996ApJ...462..547L} in analyzing the current-carrying flux emergence of AR 7265 in 1992 August. With even more advanced observations, \citet{2002ApJ...572..598K} revealed rapid changes in the $\delta$  configuration in AR 9026, that started shortly before intense flares. Based on magnetograph observations, the authors proposed a schematic cartoon containing an emerging twisted flux rope to explain the evolution of photospheric magnetic structure. Using high resolution observations at BBSO, \citet{1993Natur.363..426Z} found the new flux emergence inside the existing penumbrae of sunspots. Such an evolutional pattern produces the so-called magnetic channels, which are defined an elongated magnetic structure with alternating polarities. The strong transverse magnetic fields are also found  along the channels. Surface plasma flows along the channels are observed as well. The penumbral flux emergence were found to be a common property of  super ARs that produced multiple major flares. However, the spatial resolution of magnetograms was not sufficiently high for studying the magnetic channel structure in detail until Hinode was launched. The Solar Optical Telescope (SOT) on board Hinode provides unprecedented observations with high spatial and temporal resolutions, allowing advanced study of the nature of magnetic channels as well as their role in powering flares. \citet{2008ApJ...687..658W} and \citet{2010ApJ...719..403L} demonstrated that  high spatial resolution and high polarimetry accuracy are required to unambiguously observe the channel structure, and other complicated magnetic topology. Figure~\ref{f3} presents the magnetic channels identified by \citet{2008ApJ...687..658W} using Hinode data.

\citet{2013ApJ...764L...3C} analyzed 45~s cadence observations from SDO/HMI and reconstructed 3D subsurface magnetic structure of NOAA AR 11158. Advanced visualization methods were used in their study. The authors found that this AR consists of two major bipoles with four anchored footpoints, each of which shows tree-like structure. They concluded that an AR, even appearing highly complicated on the surface, may originate from a simple straight flux tube that undergoes both horizontal and vertical bifurcation processes during its rise through the convection zone.

On the other end, the magnetic flux cancelation is another important factor for triggering flares.  It is often closely associated with flux emergence. The importance of flux cancelation was introduced in earlier studies such as by \citet{1985AuJPh..38..929M} and \citet{1993SoPh..143..119W}. The connection between magnetic flux cancelation and flare has been established solidly in recent years using more advanced data.  e.g. \citet{2010A&A...521A..49S} used Hinode, TRACE, STEREO, and SoHO/MDI data to carry out a case study, and found strong evidence of magnetic reconnection leading to ejective eruptions. \citet{2013SoPh..283..429B} surveyed 77 X- and M-class flares,  and clearly demonstrated the importance of flux cancelation in triggering the flares.  From the aspect of the theory, a well-demonstrated example of the flux cancelation is the tether cutting model \citep{2001ApJ...552..833M}, where the first stage reconnection happens near the photosphere to form a flux rope. The erupting of the flux rope cause further reconnection in corona, leading to an ejective eruption.

\subsubsection {Flow Motions and Sunspot Rotation in Flaring Active Regions}

Evolution of magnetic field is closely associated with flow motions. For examples, photospheric flows may cause flux tubes to twist, or bring flux systems together leading to an eventual reconnection. Therefore, flow motions also play an important role in building up energy and triggering eruptions. Obviously, this motivates researchers to combine the studies of flow field and magnetic field evolution, which can provide a key piece of information for understanding the physics of flares and CMEs. In order to observe flows and magnetic field structure in detail and probe their basic nature, high-resolution and high-cadence observations are typically needed. In some studies, a simple ``center-of-mass'' (magnetic flux weighted centroids) calculation was adopted to detect the overall converging and shearing motions in the flaring ARs \citep{2005ApJ...627.1031W,2006ApJ...649..490W}. Both these flow motions show abrupt changes associated with major flares; however, no spatial information of flows is given with this method.

\citet{1976SoPh...47..233H} found strong horizontal shear motion in the chromosphere along the PILs in flaring regions. They established a strong correlation between the shear motion and flare productivity. \citet{1973SoPh...32..365M,1982PASJ...34..299M} pointed out the changes of velocity and magnetic fields, which could be associated with the onset of flares. \citet{1987A&A...185..306H} further demonstrated that flow motions are closely associated with flare productivity through the resulted rapid evolution of magnetic field, and that these motions are also related to the generation of electric currents in ARs. \citet{1993SoPh..143..107T} observed unusual flow patterns inside large $\delta$ spots. They inferred that shear motions near PILs may enhance the flare productivity of large $\delta$-spot regions. \citet{1994ASPC...68..265K} and \citet{2003A&A...412..541M} also established the linkage between shear flows and flare productivity. Nevertheless, observations of such shear motions with a subarcsecond spatial resolution are rare until recent decade when the adaptive optics and advanced data processing tools become available.

\begin{figure}
   \centering
   \includegraphics[width=10.0cm, angle=0]{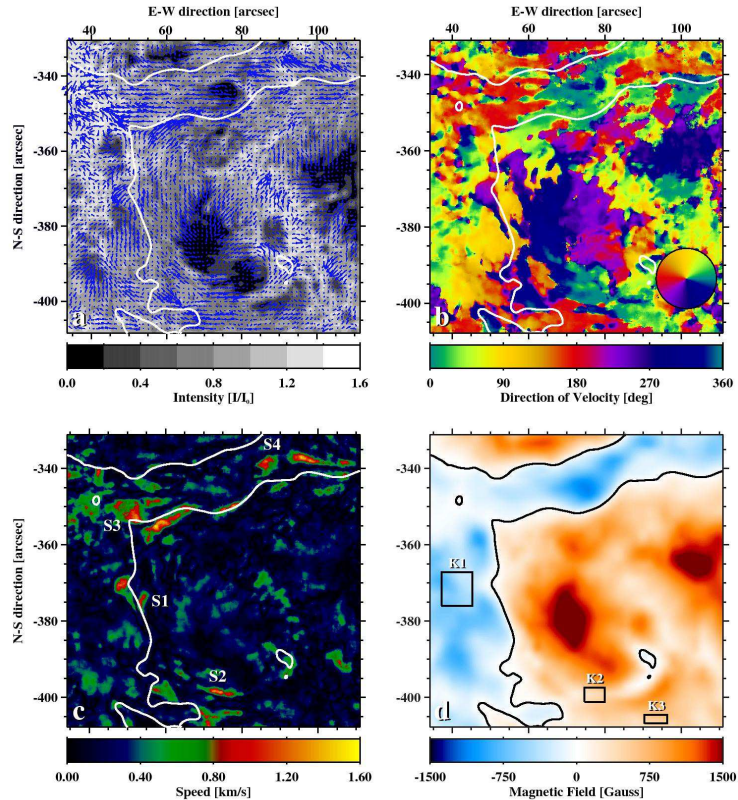}
   \caption{Photospheric flows and
    magnetic field configuration of NOAA~10486 on
    2003 November~29 \citep{2004ApJ...617L.151Y}. The authors illustrated the results of flow tracking using high spatial resolution data, and provided different views of (a) flow vectors, (b) azimuth angle of velocity vectors, (c) magnitude of velocity vectors, and (d) a corresponding LOS magnetic field image superimposed with the PIL.}
   \label{Fig4}
\end{figure}

\citet{2004ApJ...617L.151Y} presented subarcsecond resolution observations of the proper motions in AR 10486 on 2003 October 29. The data were collected at National Solar Observatory using a high-order adaptive optics system, and further processed using frame selection and speckle-masking image reconstruction. They found that flows on the two sides of the flaring PIL are almost exactly anti-parallel, which is a clear evidence of strong shear flows. These shear flows are well correlated with flare kernels in the visible and near infrared wavelengths. The maximum speed of the flow is over 1.5~km~s$^{-1}$, and the separation of channels with anti-parallel flows can be less than 1\arcsec. The authors linked the complicated flow pattern to this extremely flare-productive AR. Figure~\ref{Fig4} shows the strong shear flows observed by \citet{2004ApJ...617L.151Y}, which are clearly associated with flare kernels.

In a very recent study, using Hinode spectro-polarimeter data \citet{2014arXiv1406.1617S} found remarkable high-speed flows along the flaring PIL of the X5.4 flare on 2012 March 7. The flows lasted several hours before and after the flare. The authors argued that the observed shear flow increases the magnetic shear and free energy to power this major flare. On the theory side, \citet{2012ApJ...754...15F} simulated flux emergence and cancellation in accumulating free magnetic energy in the solar corona, and concluded that these flow motions are critical in producing solar eruptions. In short, both theory and observations point to the importance of shear flows in building up energy for solar flares.

Besides shear flows, sunspot rotation may also play an important role in the energy storage and release to power solar flares. \citet{2014ApJ...784..165R} recently found that a sunspot rotated at a high speed of 10$^{\circ}$~hr$^{-1}$. They suggested that such a sunspot rotation may cause the gradual rising of the AR filament, and therefore may be the triggering mechanism for flares. The authors also summarized the study of sunspot rotation, which was first observed by \citet{1910MNRAS..70..217E}. Many prior studies have demonstrated that the sunspot rotation is an important process in the evolution of solar active regions and triggering of eruptions \citep[e.g.,][]{1969SoPh....8..115S,1972ApJ...174..659B,1996ApJ...466L..39A,2002ApJ...567.1202T,2003A&A...406.1043T,2003SoPh..216...79B,2006A&A...451..319R,
2008MNRAS.391.1887Y,2009RAA.....9..596Y,2010ApJ...710..170S}. These studies quantitatively confirmed the important role played by sunspot rotation in transporting helicity (see next section) and energy from sub-photosphere into the corona \citep[e.g.,][]{2009ApJ...704.1146K,2012ApJ...761...60V}.  the Temporal and spatial correlation between sunspot rotation and solar eruption have been established based on careful data analysis in recent years \citep[e.g.,][]{2007ApJ...662L..35Z,2007A&A...468.1083Y,2008ApJ...682L..65Y,2008MNRAS.391.1887Y,2012ApJ...744...50J}.

\subsubsection {Magnetic Helicity and Injection}

One advanced method to analyze the interaction between the magnetic fields, flow fields, and coronal response to them is by analyzing magnetic helicity and its evolution. Magnetic helicity describes quantitatively the process that the magnetic field is sheared and twisted compared to its minimum-energy potential field topology \citep{1984JFM...147..133B,2008JApA...29...49P}. Therefore, it is  an important parameter to characterize the complexity of flaring active regions. Although the magnetic helicity is generated below the surface of the Sun, it is a useful parameter in describing magnetic structure observed above the photosphere, such as the helical patterns in filaments and CMEs, as well as spiral structure in sunspot fibrils \citep{1999GMS...111.....B}. The investigation of magnetic helicity  have been concentrated on the process of energy build-up and instability leading to flares and CMEs \citep[e.g.,][]{2001JGR...10625075R,2004ApJ...610..537K,2005ApJ...624L.129P}.\citet{2004ApJ...615.1021W} gave a comprehensive review on the relationship between the accumulation of helicity and onset of CMEs/flares.

Magnetic helicity is originally defined as the volume integral of $\textbf{A}\cdot\textbf{B}$, where \textbf{A} is the vector potential of \textbf{B}, i.e., $\nabla\times \textbf{A}=\textbf{B}$. This form is only valid for flux-enclosed systems, i.e., balanced magnetic flux on the boundary S. In such a system, the helicity is conserved. In order to calculate helicity in open systems such as the coronal magnetic fields above an active region, a modified form was introduced by adding a reference field \citep{1984JFM...147..133B}. The potential field is the most common choice of the reference field. Under this assumption, $H_r$ is written in the form by \citet{finn85}:
\begin{equation}
H_{r}=\int_{V} (\textbf{A}+\textbf{A}_{p})\cdot(\textbf{B}-\textbf{P})\,dV,
\end{equation}

\noindent where \textbf{P} is the potential field, and \textbf{A} and \textbf{A}$_p$ are vector potentials of \textbf{B} and \textbf{P}, respectively. The modification maintains its conservative property and makes the quantity gauge-invariant \citep{2008ApJ...674.1130L}. $\bf{A}$ and $\bf{A}_{p}$ can be derived following the concept of \citet{2000ApJ...539..944D} and more quantitatively by \citet{2009ApJ...697.1529F}.

As the 3D magnetic field extrapolation becomes available in recent years, calculating magnetic helicity $H_r$ in a coronal volume becomes possible.
On the other hand, because of its unique property of being conserved, one can gather information on magnetic helicity by evaluating helicity flux through the photosphere. Extensive studies of evolution of $\Delta H|_S$ have been conducted \citep{2003ApJ...594.1033N,welsch07,2010ApJ...718...43P,2010A&A...521A..56S,2011A&A...525A..13R,2011A&A...535A...1R,2011A&A...530A..36Z,2012ApJ...761...86V}.

\begin{figure}
\begin{center}
\includegraphics[scale=0.6]{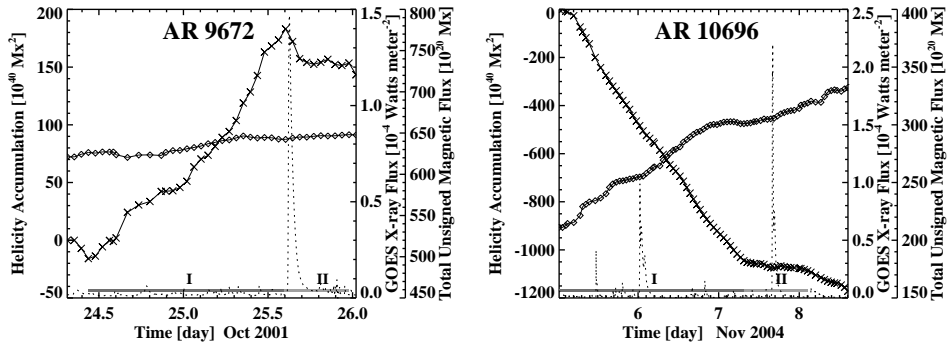}
\vskip -3mm
\caption{Time profiles of helicity accumulation, magnetic flux, and
GOES X-ray flux for two ARs. The helicity is shown as cross
symbols and the magnetic flux is shown as diamonds. The GOES X-ray
flux is shown as the dotted lines. Two phases of helicity accumulation are obvious: the preceding interval of substantial helicity accumulation (stage one) and the following phase of relatively constant helicity variation (stage two) \citep{2008ApJ...686.1397P}. \label{fig5}}
\end{center}
\end{figure}

Several studies were carried out to relate the change of magnetic helicity to the problem of impending or triggering solar flares. There are ample reports of the temporal correlation between the impulsive helicity flux changes and the occurrence of flares/CMEs \citep{2002ApJ...574.1066M,2002ApJ...580..528M,2004SoPh..223...39C,2005A&A...433..683R,2008ApJ...686.1397P,2008ApJ...682L.133Z,2010A&A...521A..56S,2011A&A...535A...1R}. Statistical studies of helicity in ARs  show that helicity in eruptive ARs is significantly higher than that in non-eruptive ones \citep{2004ApJ...616L.175N,2007ApJ...671..955L,2012ApJ...759L...4T}. In particular, it has been noticed that eruptions occur preferentially in the presence of a particular magnetic topology characterized by two magnetic flux systems with opposite helicities \citep{2003AdSpR..32.1949Y,2004ApJ...615.1021W,2004ApJ...614.1054J,2011A&A...525A..13R,2011A&A...530A..36Z}. This agrees with the MHD simulation of \citet{2004ApJ...610..537K}, in which the introduction of reversed helicity is the underlying cause of eruptions. Indeed, in a case study of the notable flare-productive AR 11158, \citet{2012ApJ...761...86V} found the sudden enhancement of reversed helicity that coincides with the onset of the X2.2 flare.

\citet{2008ApJ...686.1397P,2012ApJ...750...48P} studied the helicity flux variation for a number of events. Their results of two ARs are shown in Figure~\ref{fig5}, which demonstrates the characteristic pattern of the helicity variation found in the events analyzed by them. The helicity accumulates at a monotonic rate of change about half to two days before the flare onset, and then becomes almost constant prior to the flares. The authors concluded that typically, magnetic helicity variation has two stages:  phase I is the monotonically increasing of helicity and the phase II keeps relatively constant helicity. It was then suggested that these flares likely took place in the beginning of the phase II after a significant amount of helicity accumulation in the phase I.

\begin{figure}
\begin{center}
\includegraphics[scale=0.90]{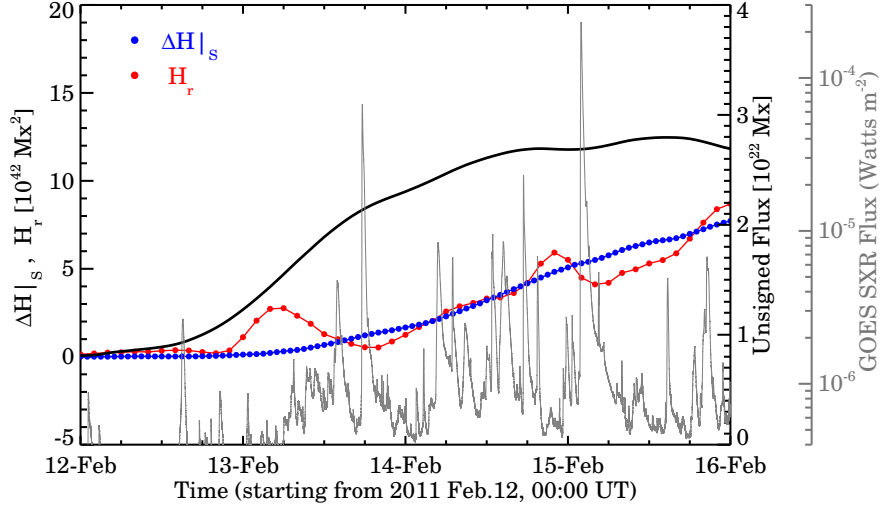}
\vskip -4mm
\caption{Temporal variation of the coronal magnetic helicity $H_r$ (red dots), the accumulated amount of helicity injection through the photosphere $\Delta H|_S$ (blue dots), and the total unsigned photospheric magnetic flux (black curve), overplotted with GOES 1--8~\AA\ soft X-ray flux (grey curve) \citep{2012ApJ...752L...9J}. \label{fig6}}
\end{center}
\end{figure}

\citet{2010ApJ...718...43P} and \citet{2012ApJ...752L...9J} compared accumulated helicity injection $\Delta H|_{S}$ (derived by integrating over time the helicity flux transported across the photosphere) with coronal magnetic helicity $H_r$ (derived by estimating helicity in a 3D volume by means of the NLFFF extrapolation). For the two ARs they studied, magnetic helicity derived from the extrapolated field is well correlated with the accumulated helicity measured from LOS magnetograms. Such an example is shown in Figure~\ref{fig6}. This gives confidence in applying NLFFF extrapolations for the 3D helicity measurement.

In order to study magnetic helicity distribution in the solar corona, \citet{2005A&A...439.1191P} proposed a new proxy to the helicity flux density that takes into account the magnetic field connectivity. Advancing in this direction, \citet{2013A&A...555L...6D,2014SoPh..289..107D} developed a method to compute such a proxy in practice. Based on analytical case studies and numerical simulations, they showed that this method can reliably and accurately determine the injection of helicity and can reveal the real mixed signals of the helicity flux.

Another useful quantity is current helicity, which is defined as the volume integral of $\textbf{B}\cdot(\nabla\times\textbf{B})$, where \textbf{B} is the magnetic field vector and $\textbf{B}\cdot(\nabla\times\textbf{B})$ is referred to as current helicity density $h_c$. The current helicity measures how much the magnetic field is locally twisted. Using the model field from NLFFF extrapolations, $h_c$ can be computed for every point in a volume; in contrast, previous works were limited to $\textbf{B}_z\cdot(\nabla\times\textbf{B})_z$ measured on the photosphere \citep{1996SoPh..168...75A,1998ApJ...496L..43B,2004PASJ...56..831H,2009ApJ...697L.103S}, or a constant force-free parameter $\alpha$ ($\nabla\times\textbf{B}=\alpha \textbf{B}$) had to be used for a whole AR \citep{1995ApJ...440L.109P}. Furthermore, vector magnetic field data from SDO/HMI allows, for the first time, the study of the evolution of $h_c$ at a cadence of 12 minute.

\begin{figure}
\begin{center}
\includegraphics[scale=0.7]{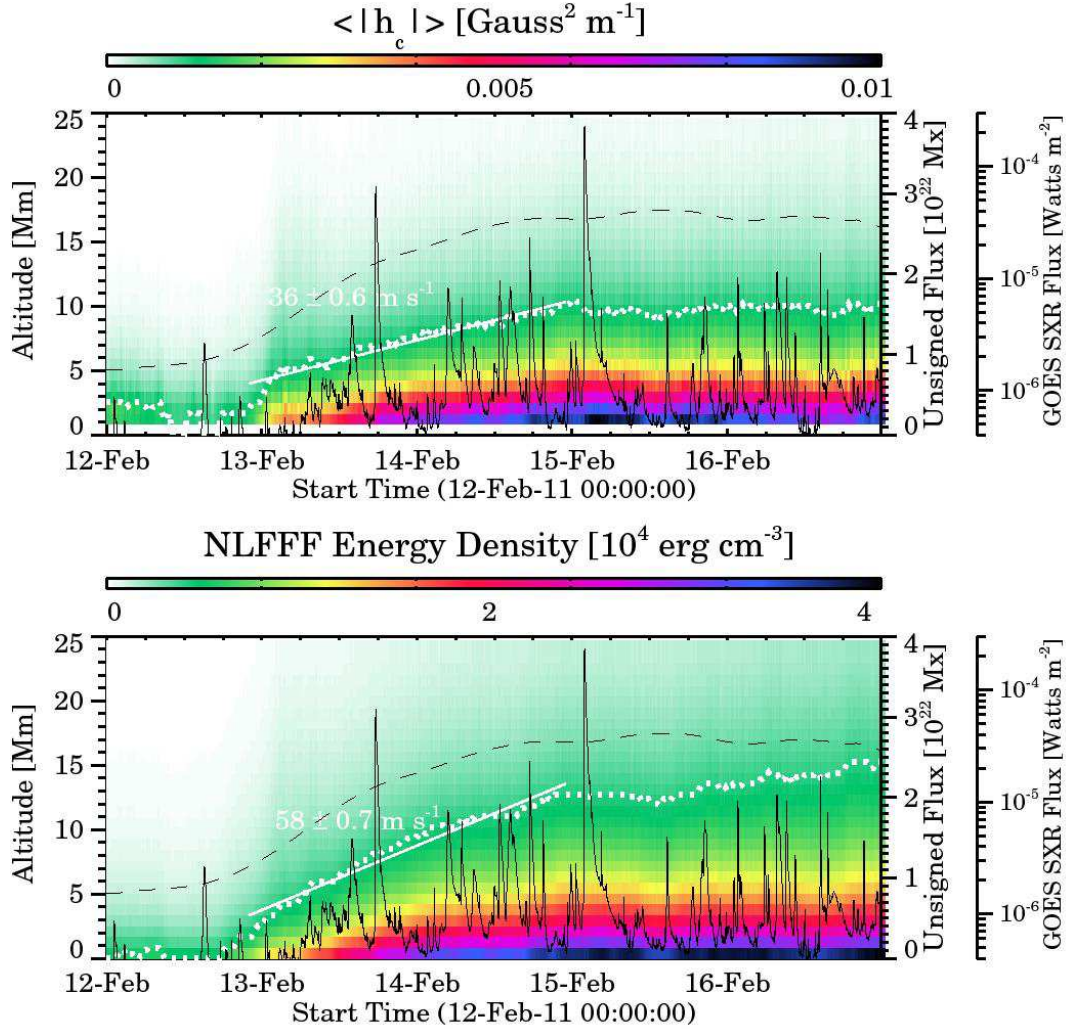}
\vskip -4mm
\caption{Time-altitude diagrams of the average unsigned current helicity density $\langle |h_c| \rangle$ (top) and average NLFFF energy density (bottom), overplotted with the time profiles of total unsigned magnetic flux (black dashed line) and GOES soft X-ray 1--8~\AA\ flux (black solid line). The white dots indicate the altitude below which the accumulated value reaches 90\%, with a linear fit for the increasing phase shown as the white solid line \citep{2012ApJ...752L...9J}. \label{fig7}}
\end{center}
\end{figure}

As shown in Figure~\ref{fig7}, \citet{2012ApJ...752L...9J} carried out NLFFF extrapolations for AR 11158 in 2011 February and found that magnetic helicity and current helicity are related to the onset of major flares. The helicity concentration also propagates to the corona as the magnetic flux emerges.

\section{Rapid Changes of Magnetic Fields Associated with Flares}
\label{sect:Change}

\subsection{Pre-SDO Research}

Although surface magnetic field evolution (such as new flux emergence and shear motions) play important roles in building energy and triggering eruptions as we discussed above, the changes on the photosphere in response to the coronal eruption are expected to be small because of the large inertia of the photosphere. One kind of reported rapid changes of the photospheric field is the so-called magnetic transients \citep{1984ApJ...280..884P,2001ApJ...550L.105K,2003ApJ...599..615Q,2009RAA.....9..812Z}, which are an apparent reversal of magnetic polarity associated with flare footpoint emissions. As the short-lived magnetic transients are regarded as an observational effect due to changes of spectral profiles, they are not considered as an rapid and irreversible change of the photospheric magnetic field associated with flares. In retrospect, \citet{1978SoPh...58..149T} detected changes of photospheric magnetic fields associated with a large flare on 1974 September 10 using the videomagnetogram at BBSO. The author explained the change as a transformation of field topology from non-potential to potential fields. However, it is unclear in this earlier study if the changes are permanent or transient. About two decades ago, BBSO group discovered obvious rapid and permanent changes of vector magnetic fields associated with flares \citep{1992SoPh..140...85W,1994ApJ...424..436W}. At that time, the authors could not find significant changes in line-of-sight magnetograms although the transverse field showed prominent changes. Part of the results appeared to be counter-intuitive: the magnetic shear angle, defined as the angular difference between the measured fields and potential fields, increases following flares. It is well known that the coronal magnetic fields have to evolve to a relaxed state to release energy in order to power flares. For this reason, there have been some doubts to these earlier measurements, especially because the data were obtained from ground-based observatories and may suffer from some seeing induced variations.

\begin{figure}
   \centering
   \includegraphics[width=11.0cm, angle=0]{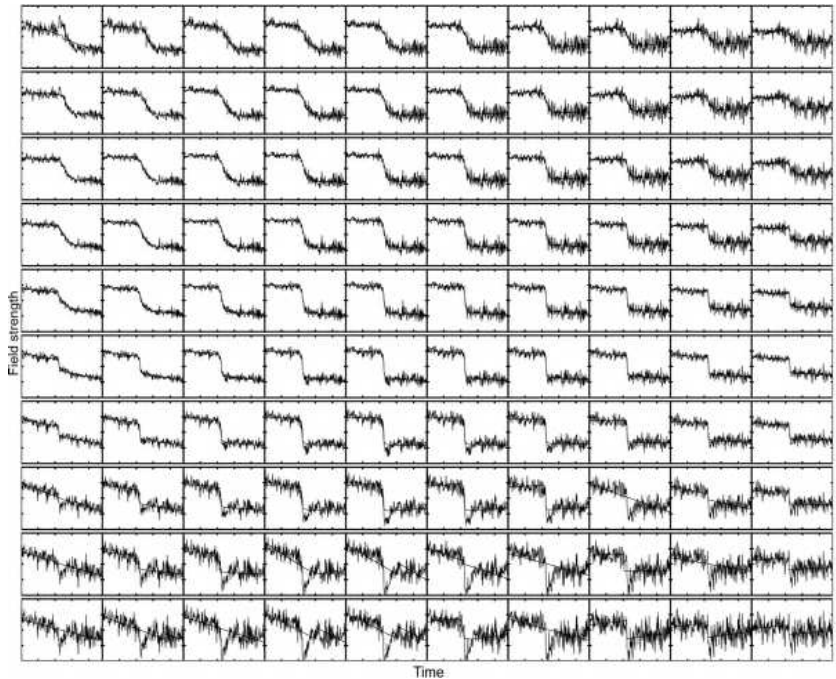}
   \caption{A mosaic of time variation plots in four hours of the longitudinal magnetic field strength, for a section of the AR that produced the flare on 2003 November 2. These plots cover a 10~$\times$~10 pixel region. The vertical axis spans 500~G. The fit using a step function is overplotted \citep{2005ApJ...635..647S}, courtesy J. Harvey.}
   \label{fig8}
   \end{figure}

\citet{2001ApJ...550L.105K} studied high-resolution MDI magnetogram data for the 2000 July 14 ``Bastille Day Flare'' and found regions with a permanent decrease of magnetic flux, which were related to the release of magnetic energy.  Using high cadence GONG data, \citet{2005ApJ...635..647S} found solid evidence of step-wise field changes associated with a number of flares.  Figure~\ref{fig8} shows the time profiles of some selected points in some selected field of view. The time scale of the changes is as fast as 10 minutes, and magnitude of change is in the order of 100~G. \citet{2010ApJ...724.1218P}, \citet{2012ApJ...760...29J}, \citet{2012ApJ...756..144C}, and \citet{2013SoPh..283..429B} also surveyed more comprehensively the rapid and permanent changes of line-of-sight magnetic fields with GONG data, which were indeed associated with almost all the X-class flares studied by them.

\begin{figure}
   \centering
   \includegraphics[width=9.0cm, angle=0]{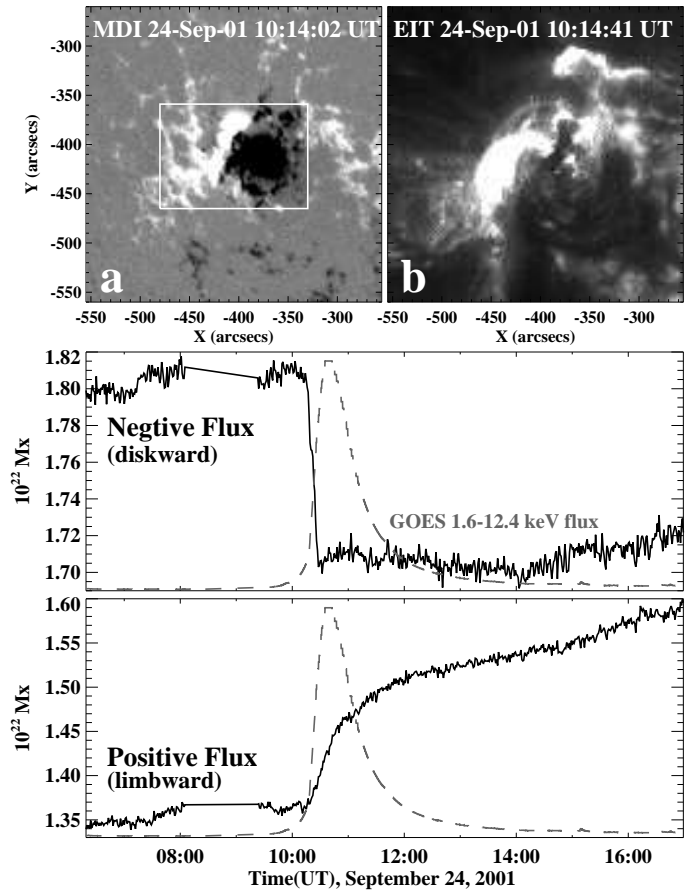}
   \caption{Time profiles of negative and positive MDI LOS magnetic fields within a boxed region (in a) covering the entire $\delta$ spot for the 2001 September 24 X2.6 flare, seen in an EUV Imaging Telescope (EIT) image (b). In GOES 10 soft X-ray flux (dashed line), the flare started at 09:32 UT, peaked at 10:38 UT, and ended at 11:09 UT \citep{2010ApJ...716L.195W}.}
   \label{Fig9}
   \end{figure}

\begin{figure}
   \centering
   \includegraphics[width=7.0cm, angle=0]{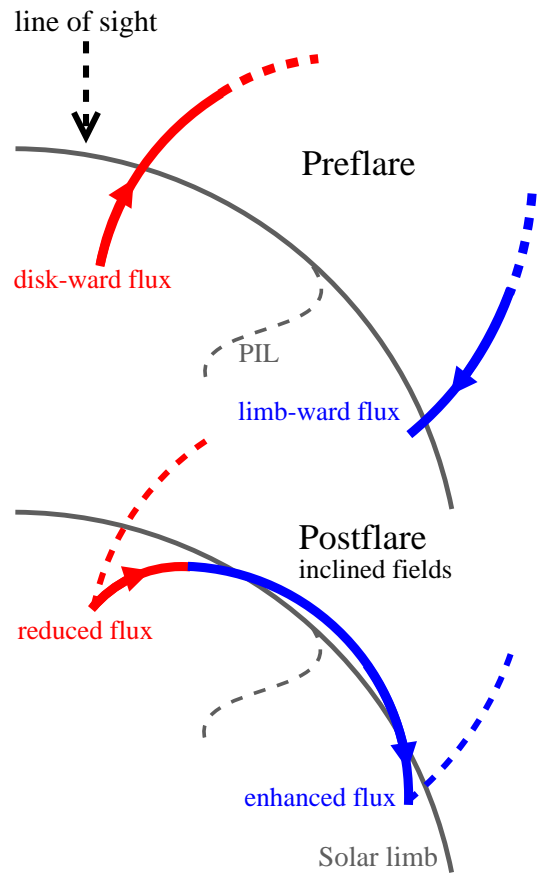}
   \caption{A conceptual illustration demonstrating the apparent changes of LOS magnetic fields when the field lines turn more horizontal with respect to the solar surface. The limbward flux would increase, while the diskward flux would decrease \citep{2010ApJ...716L.195W}.}
   \label{Fig10}
   \end{figure}

The above studies using LOS field data demonstrated the step-wise property of flare-related photospheric magnetic field change; however, the underlying physical picture was not clearly revealed. In the subsequent papers by BBSO, it was found for the LOS magnetic field that, in general, the disk-ward flux in the flaring ARs decreases while the limb-ward flux increases \citep{2002ApJ...576..497W,2006ApJ...649..490W,2004ApJ...605..546Y,2002ApJ...572.1072S,2010ApJ...716L.195W}. Such a behavior suggests that after flares, the overall magnetic field structure of ARs may change from a more vertical to a more ``flatter'' configuration, which is consistent with the scenario that Lorentz-force change pushes down the field lines (see Section~\ref{vector}). A typical example of the flare-related LOS field change is shown in Figure~\ref{Fig9} and a cartoon picture is shown in Figure~\ref{Fig10} \citep{2010ApJ...716L.195W}. Note that most of the observations listed by \citet{2010ApJ...716L.195W} are made by MDI on board SOHO, which has a cadence of one minute. It is worth mentioning that \citet{1999ApJ...525L..61C} was the first to use the near-limb magnetograph observations to characterize the flare-related changes of magnetic fields.

\begin{figure}
   \centering
   \includegraphics[width=12.0cm, angle=0]{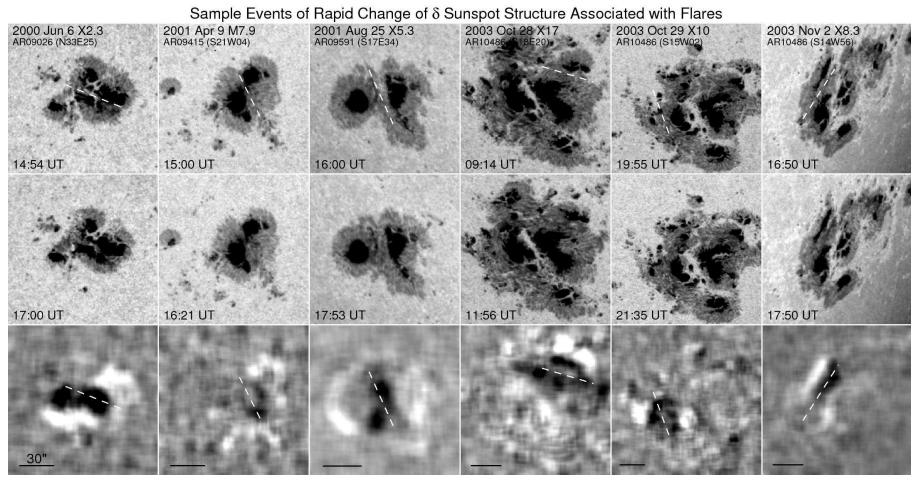}
   \caption{TRACE white-light images covering associated with six major flares. The rapid changes of $\delta$ sunspot structure are observed. The top, middle, and bottom rows
show the pre-flare, the post-flare, and the difference
images between them after some smoothing. The white pattern in the
difference image indicates the region of penumbral decay, while
the dark pattern indicates the region of darkening of penumbra.  The white dashed line denotes the flaring PIL and the black
line represents a spatial scale of 30\arcsec\ (adapted from \citet{2005ApJ...622..722L}).}
   \label{Fig11}
   \end{figure}

\begin{figure}
   \centering
   \includegraphics[width=10.0cm, angle=0]{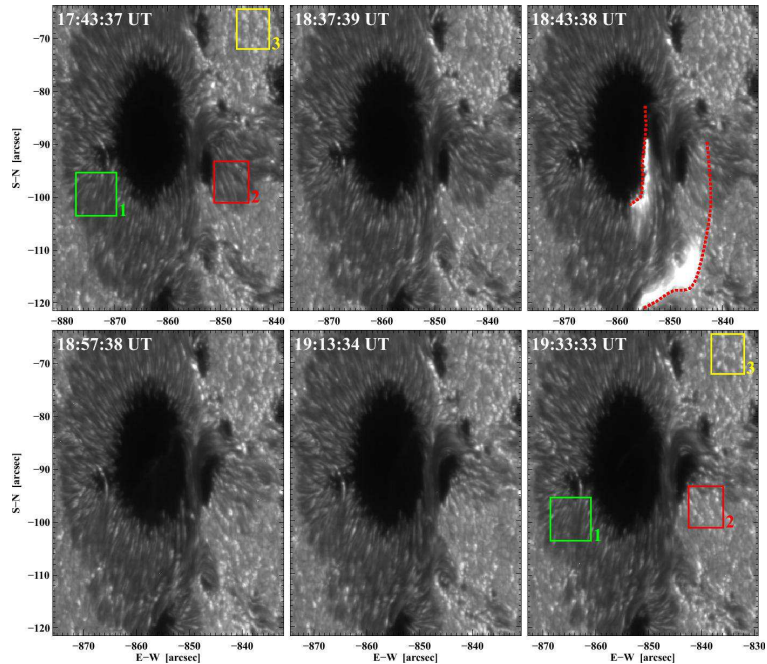}
   \caption{Time sequence of G-band images observed by Hinode/SOT on 2006 December 6
covering the X6.5 flare. The red box marks the region of a decaying penumbra (in the entire
AR 10930, there are several other penumbral decay regions). For reference, the green and yellow boxes mark the stable penumbral and facular region, respectively. The red dotted lines in the frame of 18:43:38 UT delineate the separating flare ribbons, with the western one sweeping across the penumbral decay region \citep{2012ApJ...748...76W}.}
   \label{Fig12}
   \end{figure}

\begin{figure}
   \centering
   \includegraphics[width=10.0cm, angle=0]{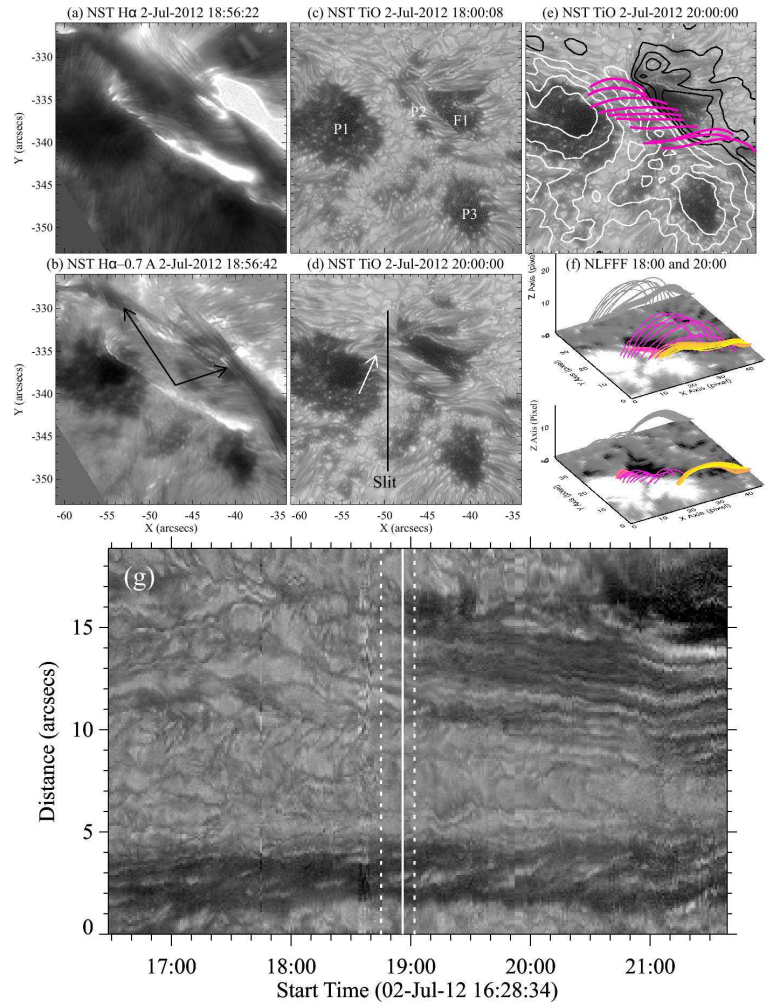}
   \vspace {-1mm}
   \caption{BBSO/NST H$\alpha$ center (a) and blue-wing (b) images at the peak of the 2011 July 2 C7.4 flare, showing the flare ribbons and possible signatures of a flux rope eruption (the arrows in (b)). The NST TiO images about 1 hr before (c) and 1 hr after (d) the flare clearly show the formation of penumbra (pointed to by the arrow in (d)). The same post-flare TiO image in (e) is superimposed with positive (white) and negative (black) HMI LOS field contours, and NLFFF lines (pink). (f) Perspective views of the pre- and post-flare 3D magnetic structure including the core field (a flux rope) and the arcade field from NLFFF extrapolations.The collapse of arcade fields is obvious \citep{2014ApJ...784L..13J}. (g) TiO time slices for a slit (black line in (d)) across the newly formed penumbra area. The dashed and solid lines denote the time of the start, peak, and end of the flare in GOES 1--8~\AA. The sudden turning off of the convection associated with the flare is obviously shown \citep{2013ApJ...774L..24W}.}
   \label{f13}
   \end{figure}

For some investigations made after 2003, researchers began to appreciate the consistent pattern of magnetic field changes associated with flares using the simple white-light observations \citep{2004ApJ...601L.195W,2005ApJ...622..722L,2005ApJ...623.1195D,2009ScChG..52.1702L,2013ApJ...774L..24W}. The most striking changes are the decay of penumbral structure in the peripheral sides of $\delta$ spots and the enhancement (darkening) of penumbral structure near the flaring PILs. Figure~\ref{Fig11} clearly demonstrates examples of such sunspot structure changes. The difference image between pre- and post-flare states always shows a dark patch at the flaring PIL that is surrounded by a bright ring. It corresponds to the enhancement of central sunspot penumbral structure and decay of the peripheral penumbrae, respectively. These examples were discussed in detail by \citet{2005ApJ...622..722L}, in which they showed that (1) these rapid changes were associated with flares and were irreversible, and (2) the decay (enhancement) of sunspot penumbrae is related to the penumbral magnetic field turning more vertical (horizontal). \citet{2007ChJAA...7..733C} statistically studied over 400 events using TRACE white-light data and found that the significance of sunspot structure change seems to be positively correlated with the magnitude of flares. Using Hinode/SOT G-band data with high spatiotemporal resolution, \citet{2012ApJ...748...76W} further characterized the penumbral decay and confirmed its intrinsic linkage to the magnetic field change. The authors took advantage of the high spatiotemporal resolution Hinode/SOT data and observed that in sections of peripheral penumbrae swept by flare ribbons, the dark fibrils completely disappear while the bright grains evolve into faculae (a signature of vertical magnetic flux tubes). These results suggest that the component of horizontal magnetic fields of the penumbra is straightened upward (i.e., turning from horizontal to vertical) due to magnetic field restructuring associated with flares. This change of magnetic topology leads to the transition of penumbrae to faculae. Figure~\ref{Fig12} shows an example of such a rapid transition of uncombed penumbral structure into faculae observed with Hinode/SOT. Also notably, the flare-related enhancement of penumbral structure near central flaring PILs has also been unambiguously observed with the 1.6~m New Solar telescope (NST) at BBSO. Using NST TiO images with unprecedented spatial (0$\farcs$1) and temporal (15~s) resolution, \citet{2013ApJ...774L..24W} reported on a rapid formation of sunspot penumbra at the PIL associated with the 2012 July 2 C7.4 flare, as presented in Figure~\ref{f13}. The most striking observation is that the solar granulation evolves to the typical pattern of penumbra consisting of alternating dark and bright fibrils. Interestingly, a new $\delta$ sunspot is created by the appearance of such a penumbra feature, and this penumbral formation also corresponds to an enhancement of the horizontal field.

\citet{2009ApJ...690..862W} carried out a detailed study of the X7.1 flare on 2005 January 20. They found clear evidence of decreasing of the horizontal magnetic fields in some peripheral areas of the $\delta$-spot group, and increasing in an extended area centered at the flaring PIL.  The observed changes are consistent with the darkening of inner penumbrae and weakening of outer penumbrae reported by other authors. The rapid magnetic changes are at the level of 100--300~G, similar to previous studies as well.

As we described earlier, for a few events studied, some authors also found that magnetic shear near the flaring PILs may increase after flares \citep[e.g.,][]{1994ApJ...424..436W,2005ApJ...622..722L}. This phenomenon was largely left unexplained, until analysis of the evolution of 3D coronal magnetic field structure associated with flares was made possible by the NLFFF extrapolation technique (see the next section).

\subsection{Observations in the SDO Era}\label{vector}

\begin{figure}
   \centering
   \includegraphics[width=12.0cm, angle=0]{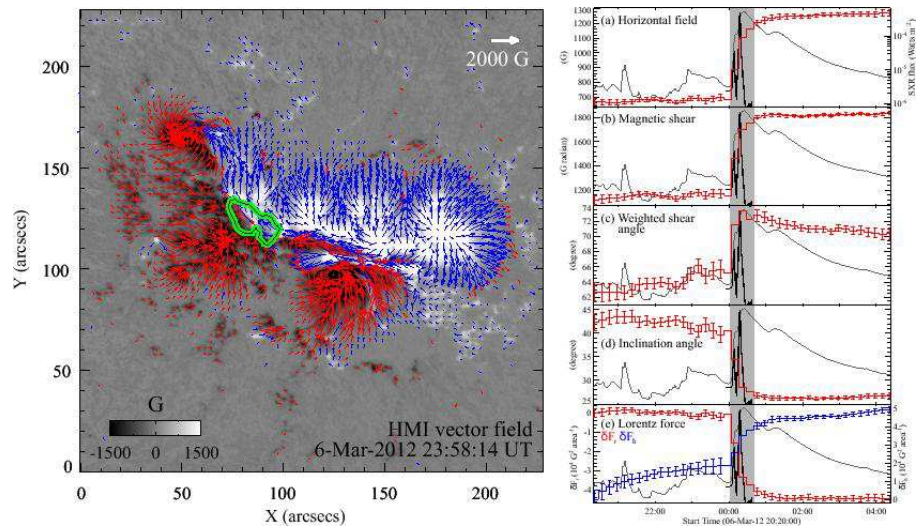}
   \caption{{\it Left}: A HMI vector magnetogram on 2012 March 7 showing the flare-productive NOAA AR 11429 right before an X5.4 flare. {\it Right}: Temporal evolution of various magnetic properties of a compact region (green contour in the left panel) at the central PIL, in comparison with the light curves of GOES 1--8~\AA\ soft X-ray flux (gray) and its derivative (black). The shaded interval denotes the flare period in GOES flux (adapted from \citet{2012ApJ...757L...5W}).}
   \label{f14}
   \end{figure}

\begin{figure}
   \centering
   \includegraphics[width=10.0cm, angle=0]{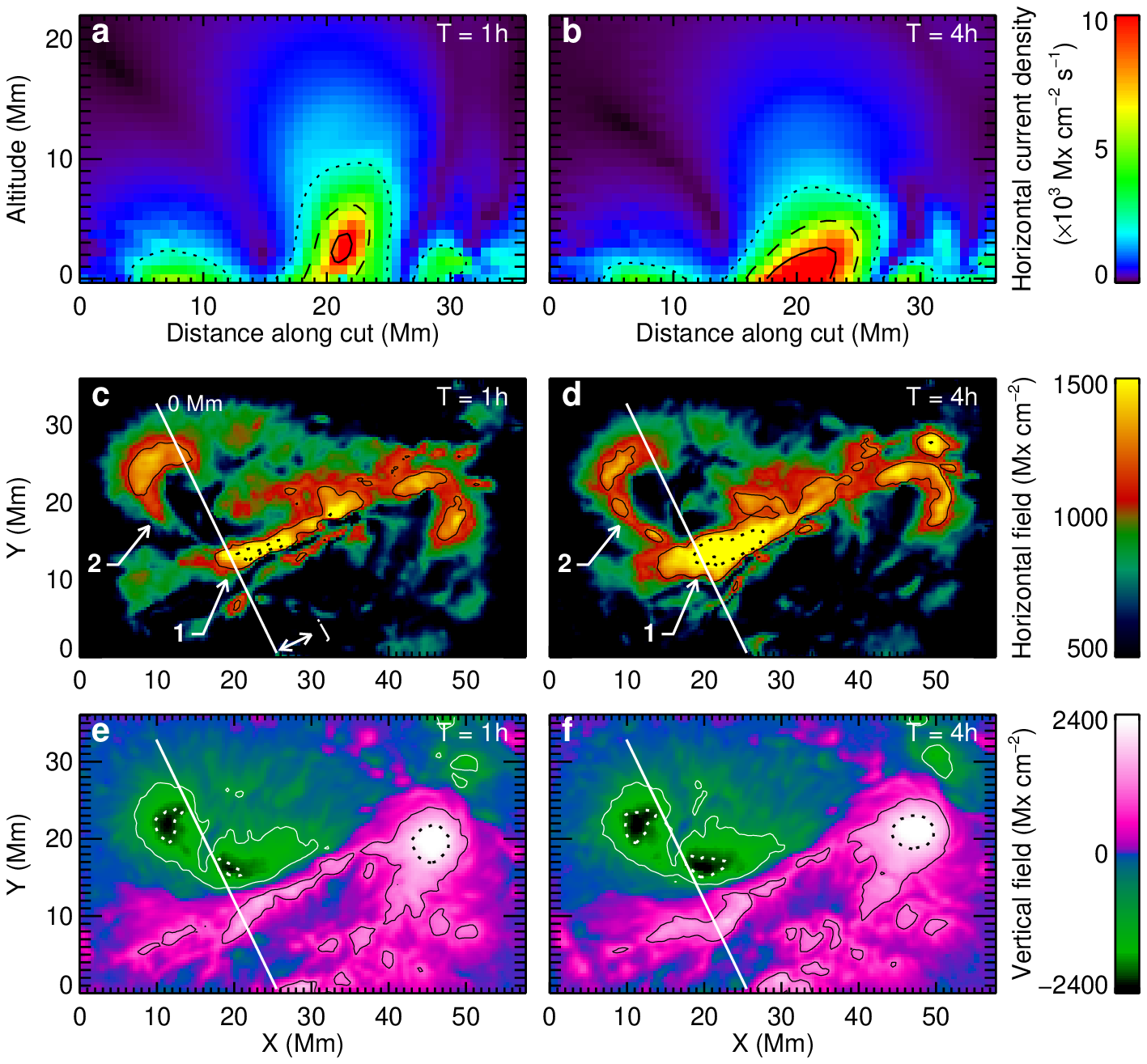}
   \caption{Modeled and observed field changes from before (01:00~UT; (a), (c), and (e)) to after (04:00~UT; (b), (d), and (f)) the 2011 February 15 X2.2 flare. (a--b) Current density distribution on a vertical cross section indicated in (c)--(f). (c--d) HMI horizontal field strength. Contour levels are 1200~G and 1500~G. (e--f) HMI vertical field. Contour levels are $\pm$1000~G and $\pm$2000~G \citep{2012ApJ...748...77S}, courtesy X. Sun.}
   \label{f15}
   \end{figure}

The launch of SDO in 2011 February provides an unprecedented opportunity to study the evolution of magnetic field associated with flares. The HMI instrument on board SDO measures vector magnetic fields and flow motions of the solar photosphere \citep{2012SoPh..275..229S}. The seeing-free, full-disk data are the key observations for many studies, as all the ARs on the visible disk are covered. The base-line HMI observations include 45~s cadence LOS magnetograms and 12 minute cadence vector magnetograms. The accuracy is in the order of 10~G for the LOS field and 100~G for the transverse field. \citet{2012ApJ...745L..17W} presented the first study of the flare-related photospheric magnetic field change using HMI data. They investigated the well-studied X2.2 flare in active region 11158 on 2011 February 15. the observations clearly demonstrated a rapid  and irreversible enhancement in the horizontal magnetic field at the flaring PIL. The mean horizonal fields increased by about 500G in half hour. The authors also found that the photospheric field becomes more sheared and more inclined. The observed field changes in this sigmoidal AR are located between the two flare ribbons, which correspond to the initial conjugate hard X-ray footpoints. RHESSI hard X-ray images and the 3D coronal magnetic field using NLFFF extrapolations are also jointly studied to understand the magnetic field changes associated with the event. These unambiguous observational evidences corroborate what were found before using ground-based observations \citep[e.g.,][]{1992SoPh..140...85W,1994ApJ...424..436W}. Later, there have been a number of papers using SDO/HMI data to demonstrate the similar changes of magnetic fields \citep{2012ApJ...745L...4L,2012ApJ...748...77S,2012ApJ...757L...5W,2012ApJ...759...50P,2013SoPh..287..415P,2014ApJ...786...72Y}. The found patterns of the changes are consistent in the sense that the transverse field enhances in a region across the central flaring PIL. Figure~\ref{f14} shows the typical time profiles of such field changes.

As we discussed in Section~\ref{sect:structure}, the NLFFF extrapolation is a powerful tool to reconstruct the 3D magnetic topology of the solar corona. Naturally, a question is raised on how the 3D magnetic field structure evolves corresponding to the observed field changes on the surface. Using Hinode/SOT magnetic field data, \citet{2008ApJ...676L..81J} showed that magnetic shear (indicating non-potentiality) only increases at lower altitude while it still largely relaxes in the higher corona. Therefore, the previously observed increase of magnetic shear along with that of the transverse field could be physically reasonable. Using HMI data \citet{2012ApJ...748...77S} clearly showed that the electric current density indeed increases at the flaring PIL near the surface while it decreases higher up, which may explain the overall decrease of free magnetic energy together with a local enhancement at low altitude (see Figure~\ref{f15}). They also found that magnetic shear increase slightly in the lower atmosphere while it relaxes rapidly in the higher atmosphere. These results imply that magnetic field could collapse toward the surface, and such a collapse was even detected in a C7.4 flare on 2012 July 2 as reported by \citet{2014ApJ...784L..13J} and shown in Figure~\ref{f13}. Intriguingly, the collapse of magnetic arcades as reflected by NLFFF models across the C7.4 flare is spatially and temporally correlated with the formation of sunspot penumbra on the surface \citep{2013ApJ...774L..24W}.

\begin{figure}
   \centering
   \includegraphics[width=11.0cm, angle=0]{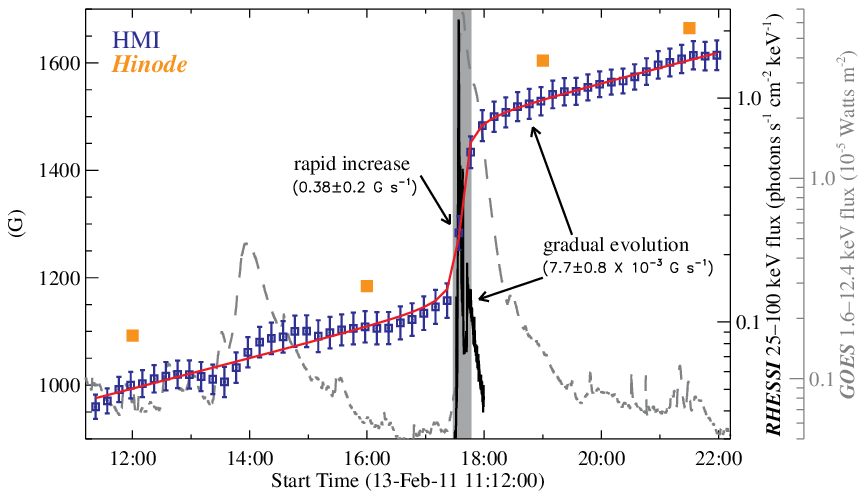}
   \includegraphics[width=11.0cm, angle=0]{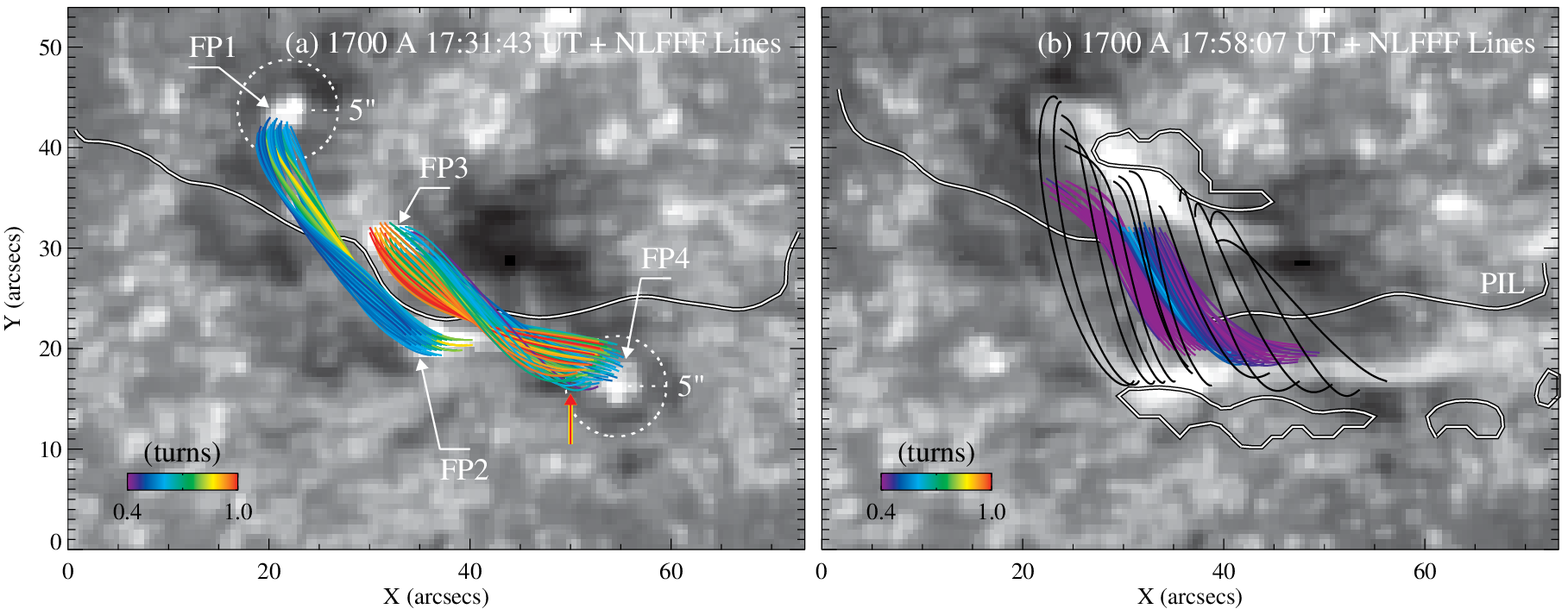}
   \caption{{\it Top}: Temporal evolution of horizontal magnetic field measured by HMI and Hinode/SOT in a compact region around the PIL, in comparison with X-ray light curves for the M6.6 flare on 2011 February 13. The red curve is the fitting of HMI data with a step function \citep{2012ApJ...745L...4L}. {\it Bottom}: Extrapolated NLFFF lines before and after the event, demonstrating the process of magnetic reconnection consistent with the tether-cutting reconnection model \citep{2013ApJ...778L..36L}.}
   \label{f16}
   \end{figure}

Using vector magnetograms from HMI together with those from Hinode/SOT with high polarization accuracy and spatial resolution, \citet{2012ApJ...745L...4L} revealed similar rapid and persistent increase of the transverse field associated with the M6.6 flare on 2011 February 13, together with collapse of coronal currents toward the surface at the sigmoid core region. \citet{2013ApJ...778L..36L} further compared the NLFFF extrapolations before and after the event. The results show that about 10\% of the flux ($\sim$3~$\times$~10$^{19}$ Mx) from the inner footpoints (e.g., FP2 and FP3 of loops FP2--FP1 and FP3--FP4) undergoes a footpoint exchange to create shorter loops of FP2--FP3. Figure~\ref{f16} presents the rapid/irreversible changes of the transverse field and the pre- and post-flare NLFFF models. These provide a direct evidence of the tether-cutting reconnection model (see the next section). A more comprehensive investigation including the 3D magnetic field restructuring, flare energy release, and also the helioseismic response, of two homologous flares, the 2011 September 6 X2.1 and September 7 X1.8 flares in NOAA AR 11283 was recently conducted by \citet{2014arXiv1409.6391L}. Their observational and model results depicted a coherent picture of coronal implosions, in which the central field collapses while the peripheral field turns vertical, as illustrated earlier by \citet{2005ApJ...622..722L}. The implosion process was also found to be more abrupt when associated with a fuller eruption.

\subsection{Theoretical Explanation of the Observations}

We realize that the physics of flares and CMEs can be ultimately
understood only if observations are coupled with theoretical
models. In particular, the afore-described observations of flare-related photospheric magnetic field changes need to be reconciled with flare/CME models. Table~\ref{Tab:publ-works} includes some of the models of interest, while it is not intended to be a complete list. Besides change of photospheric magnetic field, we also consider a number of other observational features such as remote brightenings, filament association, and sigmoid configuration \citep[e.g.,][]{2005ApJ...618.1012W,2006ApJ...649..490W,2006ApJ...642.1205L,2007ApJ...669.1372L}. These observational constraints can help assessing the applicability of different models. In a recent paper \citet{2014SoPh..289.2091L} gave a review of solar eruption models, they basically belong to 3 categories: the loss of equilibrium; tether cutting and breakout. CSHKP model is the original model for loss of equilibrium,  while erupting flux rope models are more advanced version of it.  All above models can be catastrophic.


\begin{table}
\begin{center}

\caption[]{Summary of Selected Models and Comparison with Observations. References: (1) \citet{1964NASSP..50..451C}; (2) \citet{1966Natur.211..695S}; (3) \citet{1974SoPh...34..323H}; (4) \citet{1976SoPh...50...85K}; (5) \citet{1991ApJ...373..294F};
(6) \citet{1997ApJ...490L.191C}; (7) \citet{2002ApJ...564L..53L}; (8) \citet{2006JGRA..11112103G}; (9) \citet{2010ApJ...719..728F}; (10) \citet{1980IAUS...91..207M}; (11) \citet{2001ApJ...552..833M}; (12) \citet{2006GMS...165...43M}; (13) \citet{1998ApJ...502L.181A}; (14) \citet{1999ApJ...510..485A}; (15) \citet{2004ApJ...617..589L}.}\label{Tab:publ-works}

\begin{tabular}{lcccccc}

\hline\noalign{\smallskip}

 Model & Multipolarity & Filament & CME  & B-field Change & Remote Brightening &  Reference \\

\hline\noalign{\smallskip}

Original CSHKP model      & No    &  Yes   & Yes   & No    & No   & 1,2,3,4\\
Erupting flux rope        & Maybe    &  Yes   & Yes   & Yes   & No   & 5,6,7,8,9\\
Tether-cutting            & Yes &  Yes   & Yes & Yes   & Yes  & 10,11,12\\
Breakout                  & Yes   &  Yes   & Yes   & Maybe & Yes  & 13,14,15\\

\hline\noalign{\smallskip}

\end{tabular}
\end{center}
\end{table}

\begin{figure}
   \centering
   \includegraphics[width=11.0cm, angle=0]{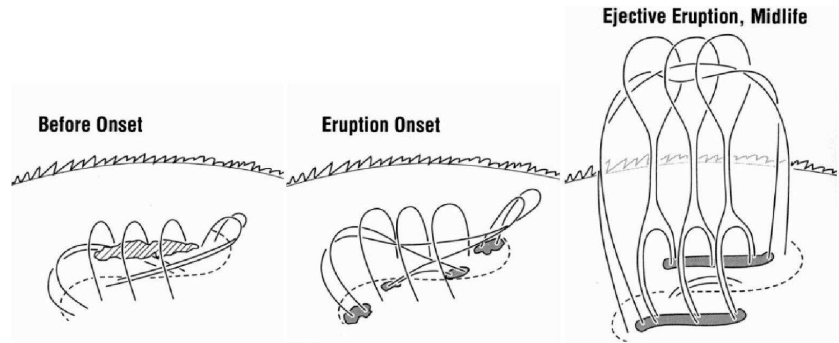}
   \caption{The tether-cutting reconnection model depicting the onset of flares and subsequent eruptions \citep{2001ApJ...552..833M}, courtesy R. Moore. There is a two-step reconnection process: the first stage is the reconnection near the solar surface, forming an erupting flux rope. The second stage involves the interaction between the erupting rope and the larger-scale arcade fields. Please note that after the eruption, a short and flat loop is formed near the photosphere, which is evidenced from our observations of enhanced transverse field near the flaring PILs and is consistent with the result of \citet{2008ASPC..383..221H}.}
   \label{f17}
   \end{figure}

All the above models can explain the observations in some ways. We use the tether-cutting model as the the example of the comparison with observations. Figure~\ref{f17} illustrates the general idea of this model as originally proposed by \citet{1980IAUS...91..207M} and further developed by \citet{2001ApJ...552..833M} and \citet{2006GMS...165...43M}.   It applies a two-step reconnection that  leads to eruptions in form of flares and CMEs, in particular, for sigmoidal active regions. The first-stage reconnection occurs at the onset of the eruption, near the solar surface. It produces a low-lying shorter loop across the PIL and a longer twisted flux rope connecting the two far ends of a sigmoid. The second stage starts when the formed twisted rope becomes unstable and erupts outward. It causes the expansion of the larger scale envelope field that overarches the sigmoid. The opened legs of the envelop fields reconnect back to form an arcade structure while the ejecting plasmoid escapes in the form of the CME. The concept of two-stage reconnection was  proposed earlier by \citet{1993SoPh..143..119W}. Please note that the tether-cutting is still  phenomenological in nature.  The full HMD modeling,  especially the data driven modeling, is required to understand the physics related to the observations.

It is possible that in the earlier phase of the flare, contraction of the shorter flare loop occurs. This has received increasing attention recently \citep[e.g.,][]{2006ApJ...636L.173J}, possibly corresponds to
the first stage. The ribbon separation as described in the standard flare models such as CSHKP model manifests the second stage. This model can also likely explain other observational findings such as: (1) transverse magnetic field at flaring PILs increases rapidly/inversibly immediately following flares \citep{2002ApJ...576..497W,2004ApJ...605..931W,2010ApJ...716L.195W}. (2) Penumbral decay occurs in the peripheral penumbral areas  of $\delta$-spots, indicating that the magnetic field lines turn more vertical after a flare in these areas \citep{2004ApJ...601L.195W,2005ApJ...622..722L}. (3) RHESSI hard X-ray images show four footpoints, two inner ones and two outer ones. Sometimes the hard X-ray emitting sources changes from  confined footpoint structure to an elongated ribbon-like structure after the flare reaches intensity maximum \citep{2007ApJ...658L.127L,2007ApJ...669.1372L}.

\begin{figure}
   \centering
   \includegraphics[width=12.0cm, angle=0]{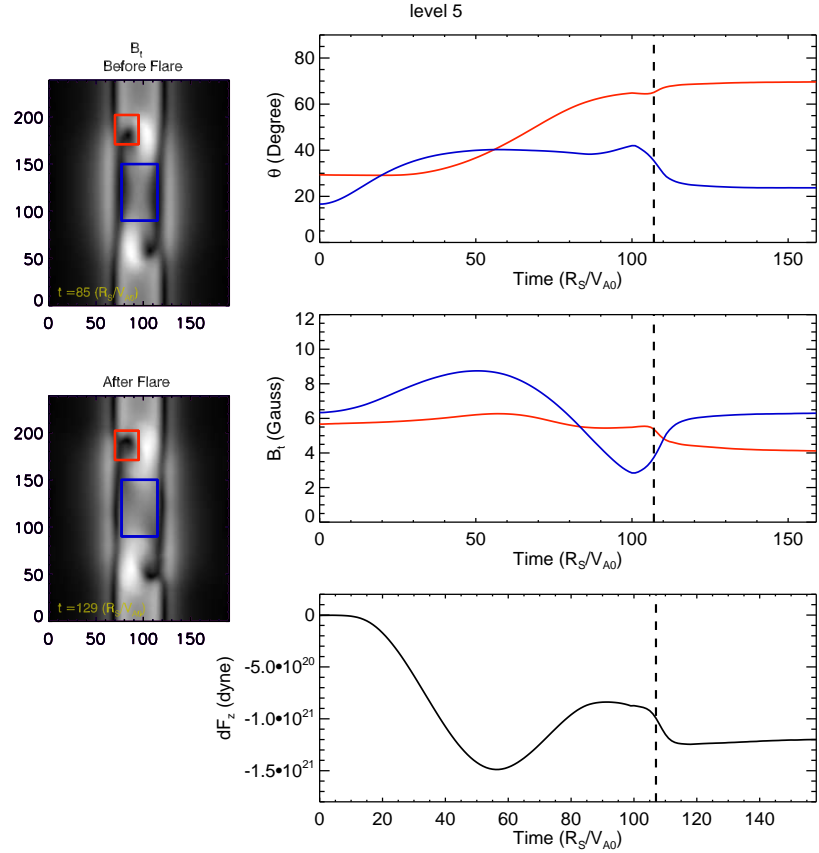}
   \caption{{\it Left panels}: The transverse magnetic field before (top) and after (bottom) the flare in $\sim$9.5~Mm above the solar surface, obtained based on the MHD simulation. The red and blue boxes denote the region of peripheral penumbral decay and darkening near the flaring PIL, respectively. {\it Right panels}: From top to bottom, temporal variation of the mean magnetic inclination angle, transverse field strength, and change of Lorentz force in the two regions illustrated in the left panels. The dashed line indicates the time of eruption. The unit of time is $R_\odot/v_{A0}=356.8$~s \citep{2011ApJ...727L..19L}.  }
   \label{f18}
   \end{figure}

An experimental comparison between simulations and observations has been made by \citet{2011ApJ...727L..19L}. Figure~\ref{f18} shows some of their results. The authors selected a lower level in the simulation to examine the near-surface magnetic structure. It is clear that the observed patterns of magnetic field change after flares/CMEs are indeed manifested by the simulation: (1) The most striking match is at the flaring PIL (blue box), where field lines in the simulation are found to be more inclined (blue line in top right panel) with a corresponding enhanced transverse field strength (blue line in middle right panel) after the eruption. Such a change of inclination angle of magnetic fields by about 10$^{\circ}$ and enhancement of transverse field were observed in several events \citep[e.g.,][]{2009ScChG..52.1702L,2010ApJ...716L.195W} and also predicted by \citet{2008ASPC..383..221H}. (2) At the outer side of the simulated sunspot penumbral area (red box), field lines turn more vertical with decreased transverse field (red lines in top and middle right panels). This strongly corroborates our previous speculation on the physical nature of the observed penumbral decay \citep{2005ApJ...622..722L}. (3) The simulation also exhibits the expected downward net Lorentz force pressing on the lower boundary (bottom right panel), which was elaborated by \citet{2008ASPC..383..221H} and \citet{2012SoPh..277...59F} and was also suggested by observations \citep{2010ApJ...716L.195W}.

The work by \citet{2008ASPC..383..221H} and \citet{2012SoPh..277...59F} has made fundamental advances in the theory to explain the observed changes of magnetic fields associated with flare. The authors quantitatively assessed the so-called back reaction on the solar surface and interior as the result of the coronal field evolution required to release energy. They made the prediction that after flares, at the flaring PIL, the photospheric magnetic fields become more horizontal. The analysis is based on the simple principle of energy and momentum conservation: the upward erupting momentum must be compensated by the downward momentum as the back reaction. This is one of the very few models that specifically predict the rapid and irreversible changes of photospheric magnetic fields associated with flares. As we discussed earlier, the tether-cutting reconnection model is closely related to this scenario. If examine the magnetic topology close to the surface, one would find a irreversible change of magnetic fields that agrees with the scenario as described above: the magnetic fields turn more horizontal near the flaring PIL. This is due to the newly formed short loops as predicted by the tether cutting model in the first stage of the reconnection.

\begin{figure}
   \centering
   \includegraphics[width=11.0cm, angle=0]{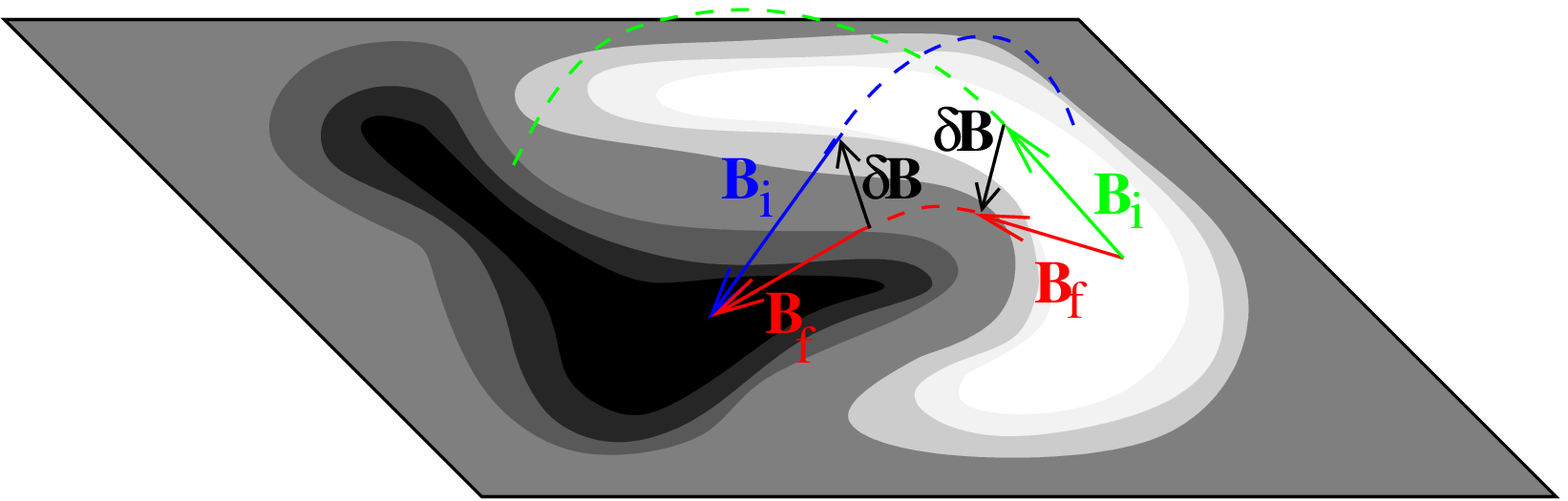}
   \caption{Model sketch showing how the initial photospheric field vector turns to a more horizontal state as a result of coronal restructuring during a flare (based upon Figure 3 of \citet{2008ASPC..383..221H}, courtesy B. Welsch). This agrees with many observational results (see text for details).}
   \label{f19}
   \end{figure}

The quantitative treatment by \citet{2008ASPC..383..221H} was carried out by \citet{2012SoPh..277...59F}, in which the changes of the integrated vertical and horizontal Lorentz force exerted on the photosphere from the corona are formulated as:

\begin{eqnarray}
\delta F_{z, {\rm downward}} = \frac{1}{8\pi}\int dA (\delta B_z^2-\delta B_h^2) \ \ {\rm and}  \\
\delta {\mathbf F_h} = \frac{1}{4\pi}\int dA \delta (B_z{\mathbf B_h}) \ ,
\end{eqnarray}

\noindent where $dA$ is the surface element of the vector magnetogram, and $B_z$ and ${\mathbf B}_h$ are the radial and horizontal components of ${\mathbf B}$. Concentrating on the downward vertical component $\delta F_z$, the authors suggested that it has to balance the upward Lorentz-force perturbation that accelerates CMEs, and they further predicted that such a change of Lorentz force might be correlated with (1) the acceleration of CMEs, and (2) the power of seismic waves excited by the downward ``jerk'' \citep{1989GMS....54..219M}. A cartoon picture demonstrating the collapse of field lines after flares is sketched in Figure~\ref{f19}.

\begin{figure}
   \centering
   \includegraphics[width=14.0cm, angle=0]{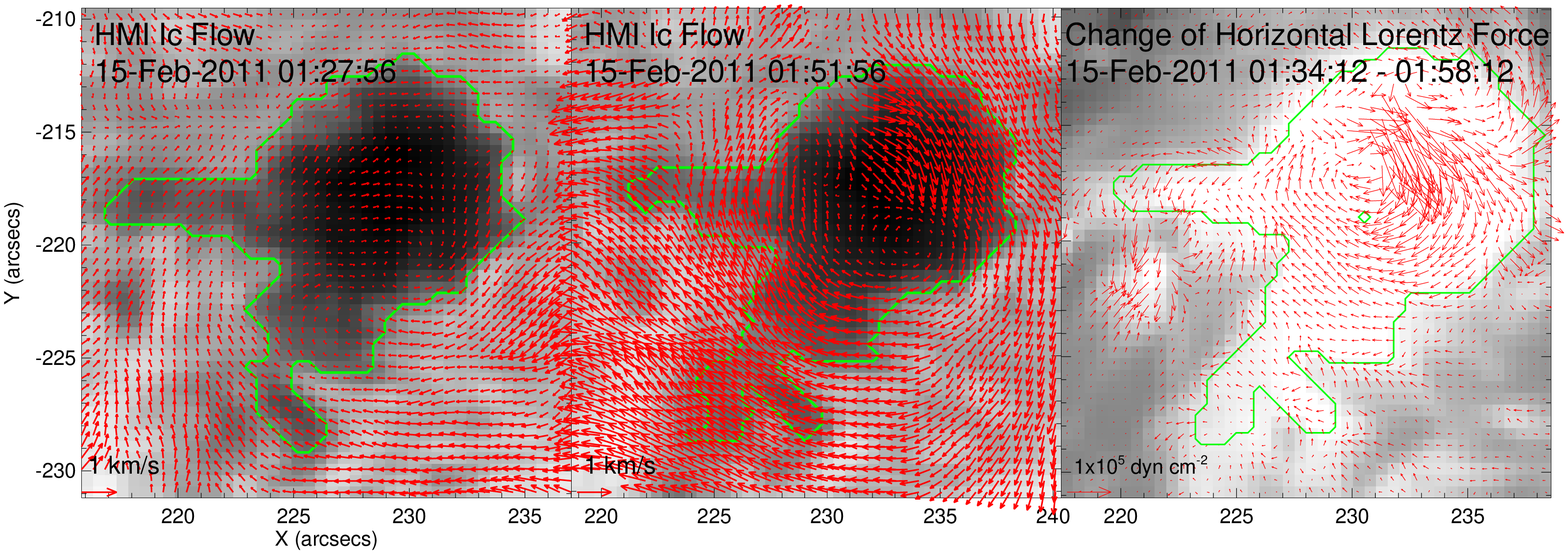}
   \includegraphics[width=14.0cm, angle=0]{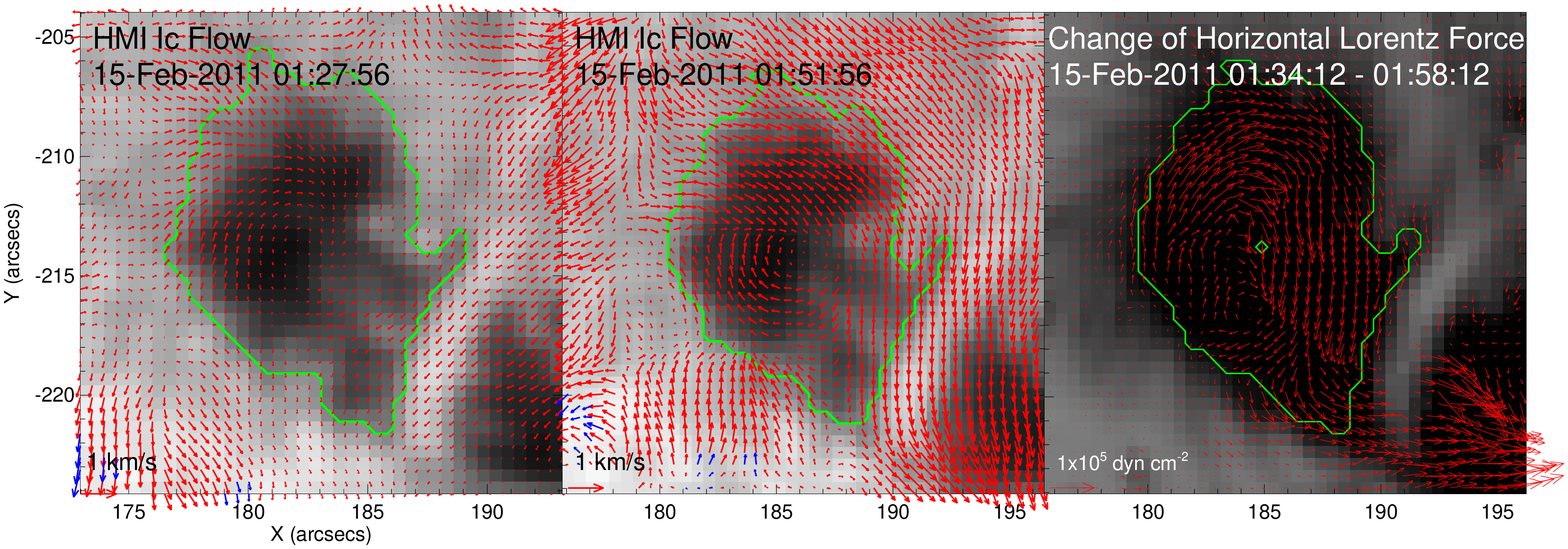}
   \caption{{\it Top}: Flow maps of the preceding spot in AR 11158 before and during the 2011 February 15 X2.2 flare and the associated change of horizontal Lorentz force. {\it Bottom}: Same as the top panels, but for the following spot in AR 11158. The direction of angular acceleration of both spots agree with the torque due to the change of horizontal Lorentz force \citep{2014ApJ...782L..31W}.}
   \label{f20}
   \end{figure}

In most of the recent studies, such a downward Lorentz force change has been detected \citep[e.g.,][]{2012ApJ...745L..17W,2012ApJ...757L...5W,2012ApJ...745L...4L,2012ApJ...748...77S,2013SoPh..287..415P}. Routine calculations of Lorentz force change is now made possible using SDO/HMI data \citep{2014SoPh..289.3663P}. Importantly, the Lorentz force perturbation as formulated above has a transverse component and may be related to sudden sunspot motion associated with flares \citep{1993SoPh..147..287A}. Very recently, \citet{2014ApJ...782L..31W} examined the photospheric sunspot structure variation after the X2.2 flare on 2011 February 15. They found that both of the two main sunspots of this AR show sudden increase of rotation speed, and that the direction of angular acceleration is consistent with the torque produced by the transverse Lorentz force change (see Figure~\ref{f20}). Under some assumptions of physical quantities, it was further shown that the amplitude of angular acceleration may agree with that derived from the torque and moment of inertia.

\section{Conclusions and Perspectives}
\label{sect:Con}

Numerous studies in the past decades have demonstrated a number of physical parameters that can reflect the non-potentiality and dynamics of flare productive solar ARs. These include, but not limited to, the magnetic free energy, magnetic shear, helicity injection, new flux emergence, shear motions, and sunspot rotation. Rapid changes of magnetic fields associated with solar flares, i.e., the central field becomes more inclined at flaring PILs while the peripheral field turns to a more vertical configuration, have also been consistently detected, not only on the photosphere but also in 3D with the aid of the NLFFF extrapolation technique. These observational and model results can be understood due to the Lorentz-force change in the framework of the back-action theory of eruptions. In particular, the tether-cutting reconnection model has been shown to be able to accommodate a various of observational features, especially the flare-related enhancement of transverse field.

It has been realized that modeling the evolution of coronal magnetic field is an important approach for gaining a comprehensive understanding of the physical nature of the flare-related magnetic field evolution. Coronal field extrapolations based on the NLFFF assumption is a useful tool, which, however, has some limitations (e.g., AR magnetic fields are not in the force-free state during the flare process). Some non-force-free modeling has been attempted. For example, \citet{2010JASTP..72..219H} developed a non-force-free extrapolation code and applied it to the real magnetic field measurement of AR 10953. The authors showed that the result was satisfactory based on quantitative evaluations. The data-driven MHD simulation may also provide a step forward in understanding the evolution of magnetic fields associated with flares \citep[e.g.,][]{2012RAA....12..563F}. \citet{2013ApJ...771L..30J} used the extrapolated NLFFF data as the initial condition for MHD simulations, which are based on the model of \cite{2006ApJ...652..800W}. They were able to achieve a realistic initiation and the subsequent evolution of the eruptive flux rope in a sigmoidal AR. The buildup process and instability condition of the flux rope was further investigated by \citet{2014ApJ...780...55J}. A comprehensive review of different models to drive solar eruptions was given by \citet{2009AdSpR..43..739S}. Obviously, a quantitative comparison between modeling results and observations is highly desirable.

Most of above discussions are associated with a common magnetic structure called sigmoidal active regions.   Flares with a closed circular-like ribbon have
been studied with recent advanced observations \citep{2009ApJ...700..559M, 2012A&A...547A..52R, 2012ApJ...760..101W, 2013ApJ...769..112D, 2013ApJ...778..139S, 2014ApJ...781L..23W}.  The events can be explained by the fan-spine magnetic topology:  the dome-shaped fan portrays the closed separatrix surface and the inner and outer spine field lines. These different connectivity domains pass through a coronal null point \citep{1990ApJ...350..672L, 2009ApJ...704..485T}. Interestingly, the outer spine can return to surface, but can also be open. In the former case, the slipping/slip-running reconnection occurs within the quasi-separatrix layers (QSLs) and leads to the sequential brightening of the circular fan ribbon. The null-point reconnection further causes the brightening in the remote area where the closed spine returns to surface \citep{2009ApJ...700..559M, 2012A&A...547A..52R}. In the latter case, the null-point reconnection surges surges/jets that erupt outward \citep{2009ApJ...691...61P, 2010ApJ...714.1762P}. These different scenarios were summarized by \citet{2012ApJ...760..101W}.  These research provide a new direction in understanding solar eruptions. The identification of skeleton structure of this kind of topology is challenging, but has been developed in recent years \citep[e.g.,][]{2008ChJAA...8..133Z,2013ApJ...778..139S}.

High-resolution observations will provide another breakthrough in detecting evolution of magnetic fields associated with flares and tracking flows in ARs. The recent completion of major telescopes such as the 1.6~m NST at BBSO, 1.5~m GREGOR at Tenerife, and 1.0~m New Vacuum Solar telescope in Yunnan, China, all demonstrate an ability to monitor the Sun with 0$\farcs$1 resolution. An example of recent flare studies using BBSO/NST images has been shown in Figure~\ref{f13}, in which the formation of sunspot penumbra is clearly observed to be associated with the occurrence of a C-class flare. Another example is the recent study by \citet{2014ApJ...781L..23W} that disclosed the fine structure of three-ribbon flares, corresponding to a complicated circular ribbon flare structure as described above. The next generation solar telescope, the 4~m Daniel K. Inouye Solar Telescope (formerly called ATST) will provide views of even finer structure of flaring ARs. More accurate polarization measurements can also be made with these large telescopes.

Finally, using helioseismology to study magnetic field evolution and the problem of flare onset will advance the physical understanding of the sub-surface magnetic and flow structures in flare productive ARs. By making an early diagnosis of ARs before they emerge to surface, helioseismology can reveal the process of energy build-up and triggering of eruptions. Studies in this direction began to produce important results in recent years. \citet{2013ApJ...762..130L} reviewed various tools for detecting pre-emerging ARs. Limitations of these tools and future studies were also discussed. The follow-up studies of this work were presented by \citet{2013ApJ...762..131B} and \citet{2014ApJ...786...19B}. Efforts have also been devoted to analyze the velocity signature of emergence process \citep[e.g.,][]{2013ApJ...777..138I} and the correlation between surface current helicity and subsurface kinetic helicity \citep{2013IAUS..294..531G}. It is anticipated that research applying helioseismic analysis to help probe the pre-flare condition will be fruitful.

\begin{acknowledgements}

We appreciate the careful reading of the manuscript and valuable comments of Prof. Jingxiu Wang. This work was supported by NSF under grants AGS 1348513 and 1408703.  Figures are largely from published papers in ApJ or ApJL, reproduced by permission of the AAS.

\end{acknowledgements}

\bibliographystyle{raa}
\bibliography{myref}


\end{document}